\documentclass[journal]{IEEEtran}

\ifCLASSINFOpdf

\else

\fi

\usepackage{latexsym}
\usepackage{amsmath,epsfig}
\usepackage{enumitem}
\usepackage{multirow}
\usepackage{subfigure}
\usepackage{multicol}
\usepackage{booktabs}
\usepackage{bigstrut}
\usepackage{array}
\usepackage{amssymb}
\usepackage{url}

\usepackage{float}
\usepackage{indentfirst}
\usepackage{times}
\usepackage{psfrag}
\usepackage{cite, xcolor}
\usepackage{textcomp}
\usepackage{multirow}
\usepackage{multicol}
\usepackage{stfloats}
\usepackage{bm, amsmath, makecell, mathrsfs}
\usepackage{hyperref}

\newcommand*{\eg}{\textit{e.g.}}
\newcommand*{\ie}{\textit{i.e.}}
\newcommand*{\etal}{\textit{et al.}}
\newcommand*{\etc}{\textit{etc.}}

\begin{document}
%
\title{Learning for Video Compression with Recurrent Auto-Encoder and Recurrent Probability Model}
%
%
%

\author{Ren~Yang,~\IEEEmembership{Student Member,~IEEE,}
        Fabian~Menzter,~\IEEEmembership{Student Member,~IEEE,} \\
        Luc~Van~Gool,~\IEEEmembership{Member,~IEEE,}
        and~Radu~Timofte,~\IEEEmembership{Member,~IEEE}
\thanks{The authors are with the Department
of Information Technology and Electrical Engineering, ETH Zurich, 8092 Zurich, Switzerland (e-mail: ren.yang@vision.ee.ethz.ch). This work is partly supported by ETH
Zurich General Fund (OK) and Amazon AWS Grant.

This paper has supplementary downloadable material available at http://ieeexplore.ieee.org, provided by the author. The material includes a supporting document. Contact ren.yang@vision.ee.ethz.ch for further questions about this work.}}

\maketitle

\begin{abstract}
The past few years have witnessed increasing interests in applying deep learning to video compression. However, the existing approaches compress a video frame with only a few number of reference frames, which limits their ability to fully exploit the temporal correlation among video frames. To overcome this shortcoming, this paper proposes a Recurrent Learned Video Compression (RLVC) approach with the Recurrent Auto-Encoder (RAE) and Recurrent Probability Model (RPM). Specifically, the RAE employs recurrent cells in both the encoder and decoder. As such, the temporal information in a large range of frames can be used for generating latent representations and reconstructing compressed outputs. Furthermore, the proposed RPM network recurrently estimates the Probability Mass Function (PMF) of the latent representation, conditioned on the distribution of previous latent representations. Due to the correlation among consecutive frames, the conditional cross entropy can be lower than the independent cross entropy, thus reducing the bit-rate. The experiments show that our approach achieves the state-of-the-art learned video compression performance in terms of both PSNR and MS-SSIM. 
Moreover, our approach outperforms the default Low-Delay P (LDP) setting of x265 on PSNR, and also has better performance on MS-SSIM than the SSIM-tuned x265 and the slowest setting of x265. The codes are available at \texttt{\url{https://github.com/RenYang-home/RLVC.git}}.
\end{abstract}


%
\IEEEpeerreviewmaketitle

\section{Introduction}\label{intro}

\IEEEPARstart{N}{owadays}, video contributes to the majority of mobile data traffic~\cite{Cisco}. The demands of high resolution and high quality video are also increasing. Therefore, video compression is essential to enable the efficient transmission of video data over the band-limited Internet. Especially, during the COVID-19 pandemic, the increasing data traffic used for video conferencing, gaming and online learning forced Netflix and YouTube to limit video quality in Europe. This further shows the essential impact of improving video compression on today's social development.

During the past decades, several video compression algorithms, such as MPEG~\cite{le1992mpeg}, H.264~\cite{wiegand2003overview} and H.265~\cite{sullivan2012overview} were standardized. These standards are handcrafted, and the modules in compression frameworks cannot be jointly optimized. Recently, inspired by the success of Deep Neural Networks (DNN) in advancing the rate-distortion performance of image compression~\cite{minnen2018joint, lee2019context,Hu2020Coarse}, many deep learning-based video compression approaches~\cite{wu2018video, lu2019dvc, cheng2019learning, djelouah2019neural, yang2020heirarchical} were proposed. In these learned video compression approaches, the whole frameworks are optimized in an end-to-end manner. 

However, both the existing handcrafted~\cite{le1992mpeg,wiegand2003overview,sullivan2012overview} and learned video compression~\cite{wu2018video, lu2019dvc, cheng2019learning, djelouah2019neural, yang2020heirarchical} approaches utilize non-recurrent structures to compress the sequential video data. As such, only a limited number of references can be used to compress new frames, thus limiting their ability for exploring temporal correlation and reducing redundancy. Adopting a recurrent compression framework enables to fully take advantage of the correlated information in consecutive frames, and thus facilitates video compression.
Moreover, in the entropy coding of previous learned approaches~\cite{wu2018video, lu2019dvc, cheng2019learning, djelouah2019neural, yang2020heirarchical}, the Probability Mass Functions (PMF) of latent representations are also independently estimated on each frame, ignoring the correlation between the latent representations among neighboring frames. Similar to the reference frames in the pixel domain, fully making use of the correlation in the latent domain benefits the compression of latent representations. Intuitively, the temporal correlation in the latent domain also can be explored in a recurrent manner. 

\begin{figure}[!t]
\centering
\includegraphics[width=.95\linewidth]{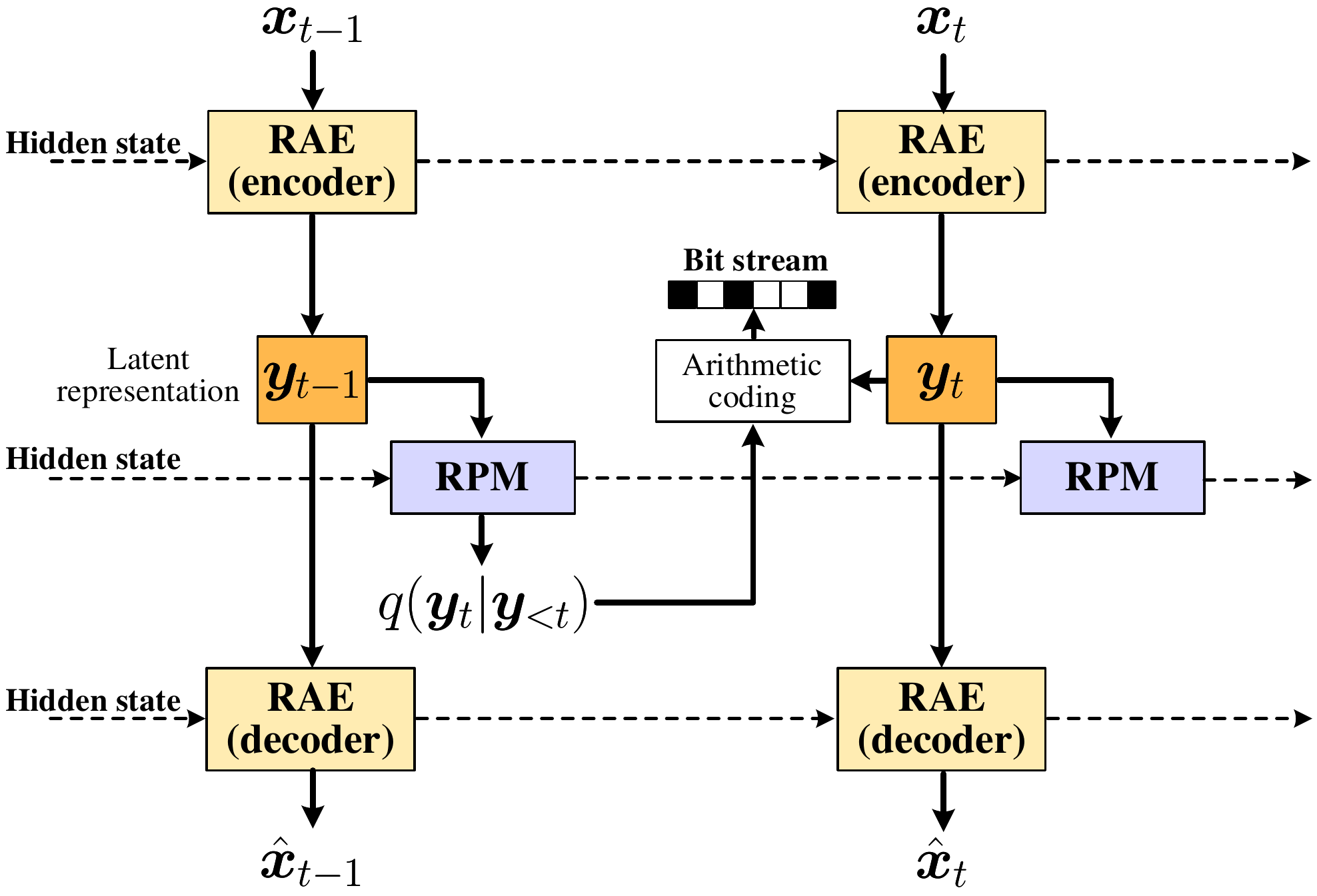}
\caption{The recurrent structure in our RLVC approach. In this figure, two time steps are shown as an example.}\label{fig:cell}
\vspace{-1em}
\end{figure}

Therefore, this paper proposes a Recurrent Learned Video Compression (RLVC) approach, with the Recurrent Auto-Encoder (RAE) and Recurrent Probability Model (RPM). 
As shown in Fig.~\ref{fig:cell}, the proposed RLVC approach uses recurrent networks for representing inputs, reconstructing compressed outputs and modeling PMFs for entropy coding.
Specifically, the proposed RAE network contains recurrent cells in both the encoder and decoder. Given a sequence of inputs $\{\bm x_t\}_{t=1}^T$, the encoder of RAE recurrently generates the latent representations $\{\bm y_t\}_{t=1}^T$, and the decoder also reconstructs the compressed outputs $\{\hat{\bm x}_t\}_{t=1}^T$ from $\{\bm y_t\}_{t=1}^T$ in a recurrent manner. As such, all previous frames can be seen as references for compressing the current one, and therefore our RLVC approach is able to make use of the information in a large number of frames, instead of the very limited reference frames in the non-recurrent approaches~\cite{wu2018video, lu2019dvc, cheng2019learning, yang2020heirarchical}. 

Furthermore, the proposed RPM network recurrently models the PMF of $\bm y_t$ conditioned on all previous latent representations $\bm y_{<t} = \{\bm y_1, \dots, \bm y_{t-1}\}$. Because of the recurrent cell, our RPM network estimates the temporally \textit{conditional} PMF $q(\bm y_t\, |\, \bm y_{<t})$, instead of the \textit{independent} PMF  $q(\bm y_t)$ as in previous works~\cite{wu2018video, lu2019dvc, cheng2019learning, djelouah2019neural, yang2020heirarchical}. Due to the temporal correlation among $\{\bm y_1, \dots, \bm y_{t}\}$, the (cross) entropy of $\bm y_t$ \textit{conditioned} on the previous information $\bm y_{<t}$ is expected to be lower than the \textit{independent} (cross) entropy. Therefore, our RPM network is able to achieve lower bit-rate to compress $\bm y_t$. As Fig.~\ref{fig:cell} illustrates, the proposed RAE and RPM networks build up a recurrent video compression framework. The hidden states for representation learning and probability modeling are recurrently transmitted from frame to frame, and therefore the information in consecutive frames can be fully exploited in both the pixel and latent domains for compressing the upcoming frames. This results in efficient video compression.

The contribution of this paper can be summarized as:
\begin{itemize}
\item We propose employing the recurrent structure in learned video compression to fully exploit the temporal correlation among a large range of video frames.
\item We propose the recurrent auto-encoder to expand the range of reference frames, and propose the recurrent probability model to recurrently estimate the temporally conditional PMF of the latent representations. This way, we achieve the expected bit-rate as the conditional cross entropy, which can be lower than the independent cross entropy in previous non-recurrent approaches.
\item The experiments validate the superior performance of the proposed approach to the existing learned video compression approaches, and the ablation studies verify the effectiveness of each recurrent component in our framework.
\end{itemize}

In the following, Section~\ref{rw} presents the related works. The proposed RAE and RPM are introduced in Section~\ref{approach}. Then, the experiments in Section~\ref{exp} validate the superior performance of the proposed RLVC approach to the existing learned video compression approaches. Finally, the ablation studies further demonstrate the effectiveness of the proposed RAE and RPM networks, respectively.

\section{Related works}\label{rw}

\textbf{Auto-encoders and RNNs.} Auto-encoders~\cite{hinton1994autoencoders} have been popularly used for representation learning in the past decades. In the field of image processing, there are plenty of auto-encoders proposed for image denoising~\cite{cho2013simple, gondara2016medical}, enhancement~\cite{lore2017llnet, park2018dual} and super resolution~\cite{zeng2015coupled, wang2016non}.
Besides, inspired by the development of Recurrent Neural Networks (RNNs) and their applications on sequential data~\cite{karpathy2015visualizing}, \eg, language modeling~\cite{mikolov2010recurrent,jozefowicz2016exploring} and video analysis~\cite{donahue2015long}, some recurrent auto-encoders were proposed for representation learning on time-series tasks, such as machine translation~\cite{cho2014learning, sutskever2014sequence} and captioning~\cite{vinyals2015show}, \etc \ Moreover, Srivastava~\etal~\cite{srivastava2015unsupervised} proposed learning for video representations using an auto-encoder based on Long Short-Term Memory (LSTM)~\cite{hochreiter1997long}, and verified the effectiveness on classification and action recognition tasks on video. However, as far as we know, there is no recurrent auto-encoder utilized in learned video compression.

\textbf{Learned image compression.} In recent years, there are increasing interests in applying deep auto-encoders in the end-to-end DNN models for learned image compression~\cite{Toderici2016Variable,toderici2017full,agustsson2017soft,theis2017lossy, balle2017end,balle2018variational,minnen2018joint,mentzer2018conditional,li2018learning,johnston2018improved, lee2019context,Hu2020Coarse}. For instance, Theis~\etal~\cite{theis2017lossy} proposed a compressive auto-encoder for lossy image compression, and reached competitive performance with JPEG 2000~\cite{skodras2001jpeg}. Later, various probability models were proposed. For instance,
Ball{\'e}~\etal~\cite{balle2017end, balle2018variational} proposed the factorized prior~\cite{balle2017end} and hyperprior~\cite{balle2018variational} probability models to estimate entropy in the end-to-end DNN image compression frameworks. Later, based on them, Minnen~\etal~\cite{minnen2018joint} proposed the hierarchical prior entropy model to improve the compression efficiency. 
Besides, Mentzer~\etal~\cite{mentzer2018conditional} utilized 3D-CNN as the context model for entropy coding, and proposed learning an importance mask to reduce the redundancy in latent representation. 
Recently, the  context-adaptive~\cite{lee2019context} and the coarse-to-fine hyper-prior~\cite{Hu2020Coarse} entropy models were designed to further advance the rate-distortion performance, and successfully outperform the traditional image codec BPG~\cite{BPG}.

\textbf{Learned video compression.} 
Deep learning is also attracting more and more attention in video compression. To improve the coding efficiency of handcrafted standard, many approaches~\cite{xu2018reducing, liu2018one, choi2019deep, dai2017convolutional, li2019densenet, li2019deep} were proposed to replace the components in H.265 by DNN. Among them, Liu~\etal~\cite{liu2018one} utilized DNN in the fractional interpolation of motion compensation, and Choi~\etal~\cite{choi2019deep} proposed a DNN model for frame prediction. Besides, \cite{dai2017convolutional, li2019densenet, li2019deep} employed DNNs to improve the in-loop filter of H.265. However, these approaches only advance the performance of one particular module, and the video compression frameworks cannot be jointly optimized.

Inspired by the success of learned image compression, some learning-based video compression approaches were proposed \cite{chen2017deepcoder,chen2019learning}. However, \cite{chen2017deepcoder,chen2019learning} still adopt some handcrafted strategies, such as block matching for motion estimation and compensation, and therefore they fail to optimize the whole compression framework in an end-to-end manner.
Recently, several end-to-end DNN frameworks have been proposed for video compression~\cite{wu2018video, lu2019dvc, cheng2019learning, djelouah2019neural, habibian2019video, liu2019learned, yang2020heirarchical}. Specifically, Wu~\etal~\cite{wu2018video} proposed predicting frames by interpolation from reference frames, and compressing residual by the image compression model~\cite{toderici2017full}. Later, Lu~\etal~\cite{lu2019dvc} proposed the Deep Video Compression (DVC) approach, which uses optical flow for motion estimation, and utilizes two auto-encoders to compress the motion and residual, respectively. Then, Djelouah~\etal~\cite{djelouah2019neural} employs bi-directional prediction in to learned video compression. Liu~\etal~\cite{liu2019learned} proposed a deep video compression framework with the one-stage flow for motion compensation. Most recently, Yang~\etal~\cite{yang2020heirarchical} proposed learning for video compression with hierarchical quality layers and adopted a recurrent enhancement network in the deep decoder. Agustsson~\etal~\cite{agustsson2020scale} proposed the scale-space flow for learned video compression, which learns to adaptively blur frame if the bilinearly warped frame is not a good prediction. Nevertheless, none of them learns to compress video with a recurrent model. Instead, there are at most two reference frames used in these approaches~\cite{wu2018video, lu2019dvc, cheng2019learning, djelouah2019neural, liu2019learned, yang2020heirarchical}, and therefore they fail to exploit the temporal correlation in a large number of frames. 

Although Habibian~\etal~\cite{habibian2019video} proposed taking a group of frames as inputs to the 3D auto-encoder, the temporal length is limited as all frames in one group have to fit into GPU memory \emph{at the same time}. Instead, the proposed RLVC network takes as inputs only one frame and the hidden states from the previous frame, and recurrently moves forward. Therefore, we are able to explore larger range of temporal correlation with finite memory. Also, \cite{habibian2019video} uses a PixelCNN-like network~\cite{van2016conditional} as an auto-regressive probability model, which makes decoding slow. On the contrary, the proposed RPM network benefits our approach to achieve not only more efficient compression but also faster decoding.   

\section{The proposed RLVC approach}\label{approach}

\begin{figure*}[!t]
\centering
\includegraphics[width=.9\linewidth]{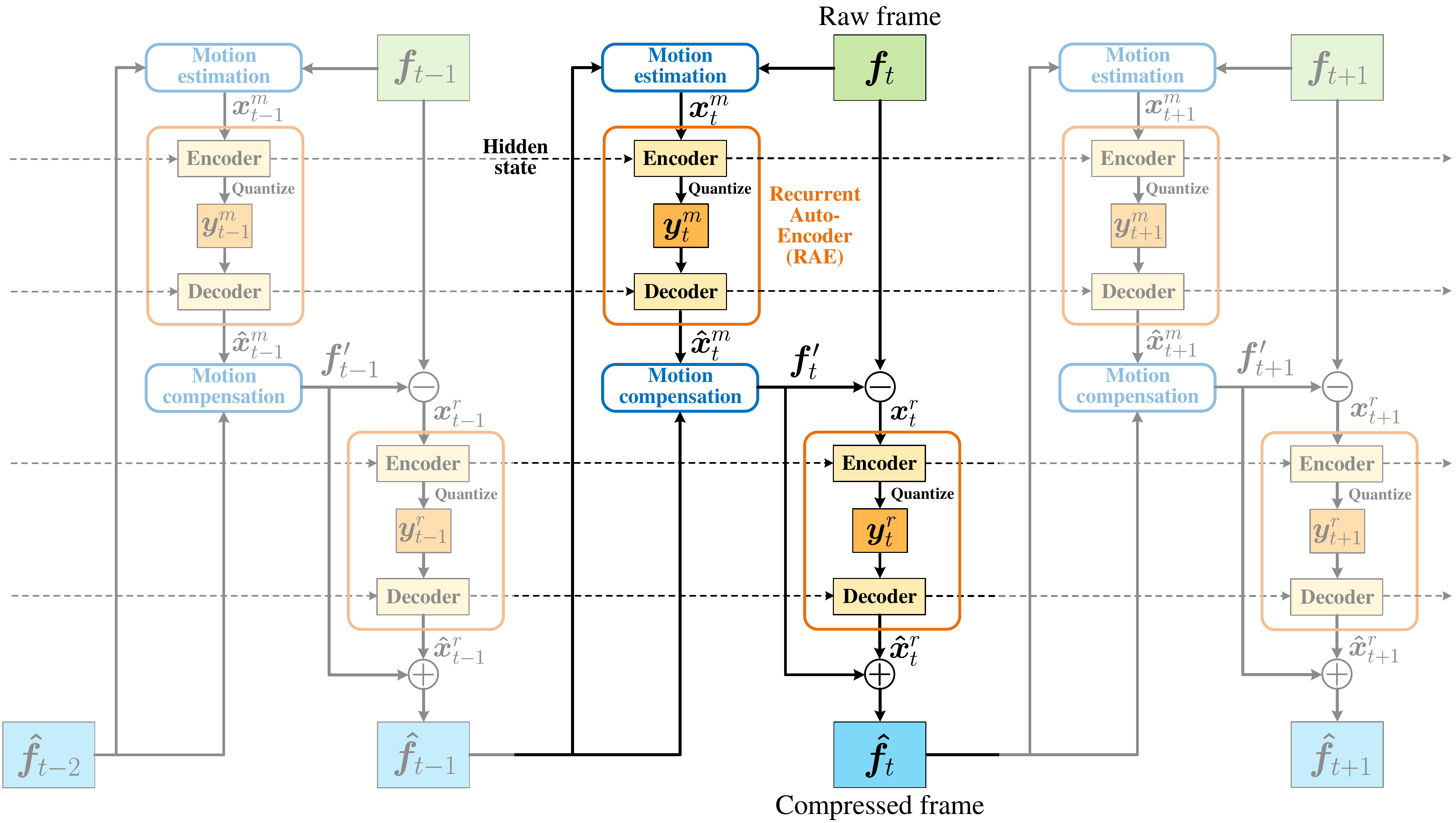}
\caption{The framework of our RLVC approach. The details of the proposed RAE are shown in Fig.~\ref{fig:rae}. The proposed RPM, which is illustrated in Fig.~\ref{fig:RPM}, is applied on the latent representations $\bm{y}^m_t$ and $\bm{y}^r_t$ to estimate their conditional PMF for arithmetic coding.} \label{fig:framework}
\end{figure*}

\subsection{Framework}

The framework of the proposed RLVC approach is shown in Fig.~\ref{fig:framework}. Inspired by traditional video codecs, we utilize motion compensation to reduce the redundancy among video frames, whose effectiveness in learned compression has been proved in previous works \cite{lu2019dvc, yang2020heirarchical}. To be specific, we apply the pyramid optical flow network~\cite{ranjan2017optical} to estimate the temporal motion between the current frame and the previously compressed frame, \eg, $\bm{f}_t$ and $\bm{\hat{f}}_{t-1}$. The large receptive field of the pyramid network~\cite{ranjan2017optical} benefits to handle large and fast motions. Here, we define the raw and compressed frames as $\{\bm{f}_t\}_{t=1}^T$ and $\{\bm{\hat{f}}_t\}_{t=1}^T$, respectively. Then, the estimated motion $\bm{x}^m_t$ is compressed by the proposed RAE, and the compressed motion $\bm{\hat{x}}^m_t$ is applied for motion compensation. In our framework, we use the same motion compensation method as \cite{lu2019dvc, yang2020opendvc, yang2020heirarchical}. 
In the following, the residual ($\bm{x}^r_t$) between $\bm{f}_t$ and the motion compensated frame $\bm{f}'_t$ can be obtained and compressed by another RAE. Given the compressed residual as $\bm{\hat{x}}^r_t$, the compressed frame $\bm{\hat{f}}_t = \bm{f}'_t + \bm{\hat{x}}^r_t$ can be reconstructed. The details of the proposed RAE is described in Section~\ref{rae_sec}.

In our framework, the two RAEs in each frame generate the latent representations of $\bm{y}^m_t$ and $\bm{y}^r_t$ for motion and residual compression, respectively. To compress $\bm{y}^m_t$ and $\bm{y}^r_t$ into a bit stream, we propose the RPM network to recurrently predict the temporally \textit{conditional} PMFs of $\{\bm{y}^m_t\}_{t=1}^{T}$ and $\{\bm{y}^r_t\}_{t=1}^{T}$. Due to the temporal relationship among video frames, the \textit{conditional} cross entropy is expected to be lower than the \textit{independent} cross entropy used in non-recurrent approaches \cite{wu2018video, lu2019dvc, cheng2019learning, yang2020heirarchical}. Hence, utilizing the conditional PMF estimated by our RPM network effectively reduces bit-rate in arithmetic coding~\cite{langdon1984introduction}.
The proposed RPM is detailed in Section~\ref{RPM}.

\begin{figure*}[!t]
\centering
\includegraphics[width=.9\linewidth]{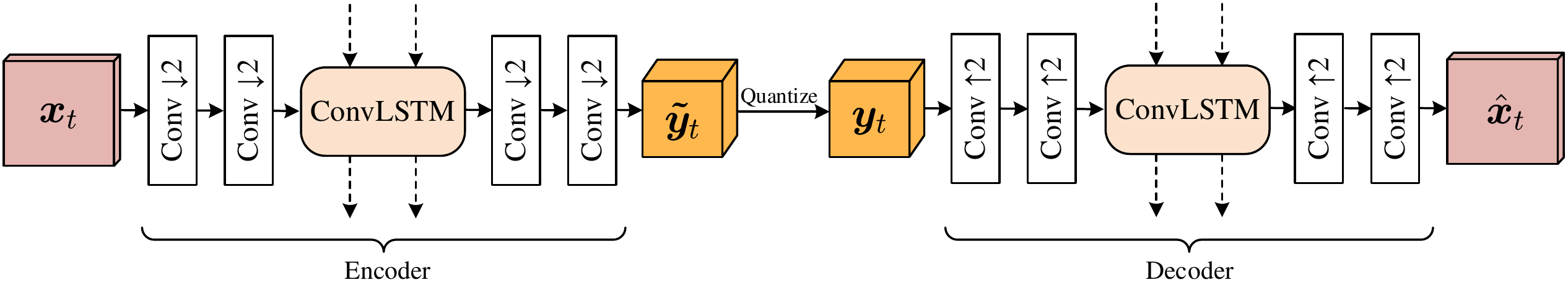}
\caption{The architecture of the proposed RAE network. In convolutions layers, $\uparrow 2$ and $\downarrow 2$ indicate up- and down-sampling with the stride of 2, respectively. In RAE, the filter sizes of all convolutional layers are set as $3\times3$ when compressing motion, and set as $5\times5$ for residual compression. The filter number of each layer is set as 128.} \label{fig:rae}
\end{figure*}

\subsection{Recurrent Auto-Encoder (RAE)}\label{rae_sec}

As mentioned above, we apply two RAEs to compress $\bm{x}^m_t$ and $\bm{x}^r_t$. Since the two RAEs share the same architecture, we denote both $\bm{x}^m_t$ and $\bm{x}^r_t$ by $\bm{x}_t$ in this section for simplicity. 
Recall that in the non-recurrent learned video compression works~\cite{lu2019dvc, cheng2019learning, yang2020heirarchical}, when compressing the $t$-th frame, the auto-encoders map the input $\bm{x}_t$
to a latent representation 
\begin{equation}
    \bm{\tilde{y}}_t = E(\bm{x}_t; \bm \theta_E)
\end{equation}
through an encoder $E$ parametrized with $\bm \theta_E$. Then, the continuous-valued $\bm{\tilde{y}}_t$ is quantized to the discrete-valued $\bm y_t=\lfloor \bm{\tilde{y}}_t \rceil$. The compressed output is reconstructed by the decoder from the quantized latent representation, \ie, 
\begin{equation}
    \hat{\bm{x}}_t = D(\bm{y}_t; \bm \theta_D).
\end{equation}
Taking the inputs of only the current $\bm x_t$ and $\bm y_t$ to the encoder and decoder, they fail to take advantage of the temporal correlation in consecutive frames. 

On the contrary, the proposed RAE includes recurrent cells in both the encoder and decoder. The architecture of the RAE network is illustrated in Fig.~\ref{fig:rae}. We follow~\cite{balle2018variational} to use four $2{\times}$ down-sampling convolutional layers with the activation function of GDN~\cite{balle2017end} in the encoder of RAE. In the middle of the four convolutional layers, we insert a ConvLSTM~\cite{xingjian2015convolutional} cell to achieve the recurrent structure. As such, the information from previous frames flows into the encoder network of the current frame through the hidden states of the ConvLSTM. Therefore, the proposed RAE generates latent representation based on the current \emph{as well as} previous inputs. Similarly, the recurrent decoder in RAE also has a ConvLSTM cell in middle of the four $2\times$ up-sampling convolutional layers with IGDN~\cite{balle2017end}, and thus also reconstructs $\bm{\hat{x}}_t$ from both the current and previous latent representations. In summary, our RAE network can be formulated as
\begin{equation} \label{rae}
    \begin{aligned}
    \bm y_t &= \lfloor E(\bm x_1, \dots, \bm x_t; \bm \theta_E)\rceil, \\
    \bm{\hat{x}}_t &= D(\bm y_1, \dots, \bm y_t; \bm \theta_D).
    \end{aligned}
\end{equation}
In \eqref{rae}, all previous frames can be seen as reference frames for compressing the current frame, and therefore our RLVC approach is able to make use of the information in a large range of frames, instead of the very limited number of reference frames in the non-recurrent approaches~\cite{wu2018video, lu2019dvc, cheng2019learning, yang2020heirarchical}. 

\subsection{Recurrent Probability Model (RPM)}\label{RPM}

To compress the sequence of latent representations $\{\bm y_t\}_{t=1}^T$, the RPM network is proposed for entropy coding. First, we use $p(\bm y_t)$ and $q(\bm y_t)$ to denote the true and estimated \textit{independent} PMFs of $\bm y_t$. The expected bit-rate of $\bm y_t$ is then given as the cross entropy
\begin{equation}\label{CE}
H(p, q) = 
\mathbb{E}_{\bm y_t \sim p}[-\log_2 q(\bm y_t)].   
\end{equation}
Note that arithmetic coding~\cite{langdon1984introduction} is able to encode $\bm y_t$ at the bit-rate of the cross entropy with negligible overhead.
It can be seen from \eqref{CE} that if $\bm y_t$ has higher certainty, 
the bit-rate can be smaller. 
Due to the temporal relationship among video frames, the distribution of $\bm y_t$ in consecutive frames are correlated. Therefore, conditioned on the information of previous latent representations $\bm y_1, \dots, \bm y_{t-1}$, the current $\bm y_t$ is expected to be more certain. That is, defining $p_t(\bm y_t\, | \,\bm y_1, \dots, \bm y_{t-1})$ and $q_t(\bm y_t\, | \,\bm y_1, \dots, \bm y_{t-1})$ as the true and estimated temporally \textit{conditional} PMF of $\bm y_t$, the \textit{conditional} cross entropy
\begin{equation}\label{CE_new}
H(p_t, q_t) = 
\mathbb{E}_{\bm y_t \sim p_t}[-\log_2 q_t(\bm y_t\, | \,\bm y_1, \dots, \bm y_{t-1})]   
\end{equation}
can be smaller than the \textit{independent} cross entropy in \eqref{CE}. To achieve the expected bit-rate of \eqref{CE_new}, we propose the RPM network to recurrently model the \textit{conditional} PMF $q_t(\bm y_t\, | \,\bm y_1, \dots, \bm y_{t-1})$.

\begin{figure}[!t]
\centering
\includegraphics[width=.55\linewidth]{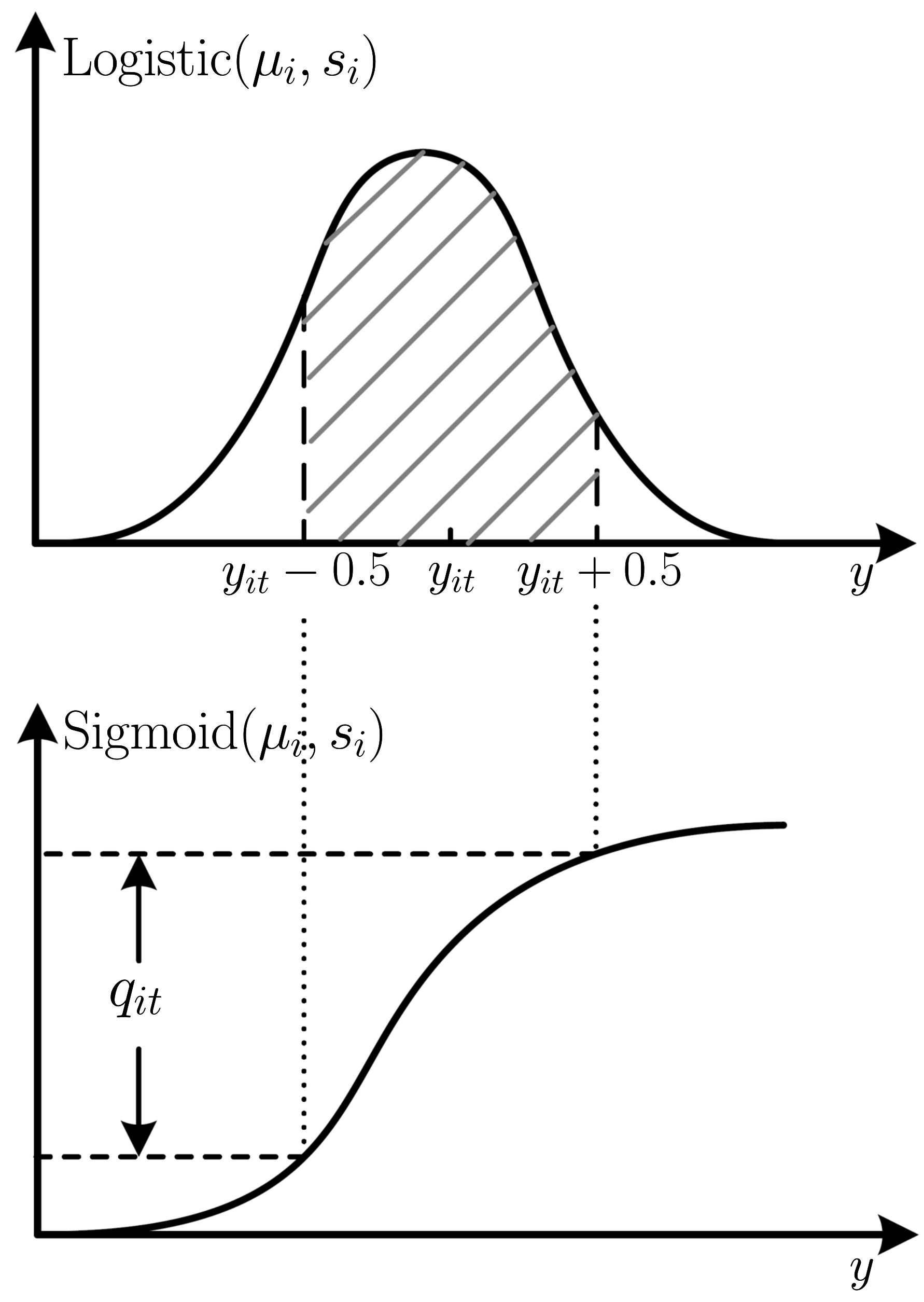}
\caption{Modeling the conditional PMF with a discretized logistic distribution.}\label{fig:log}
\end{figure}

\begin{figure}[!t]
\centering
\includegraphics[width=.99\linewidth]{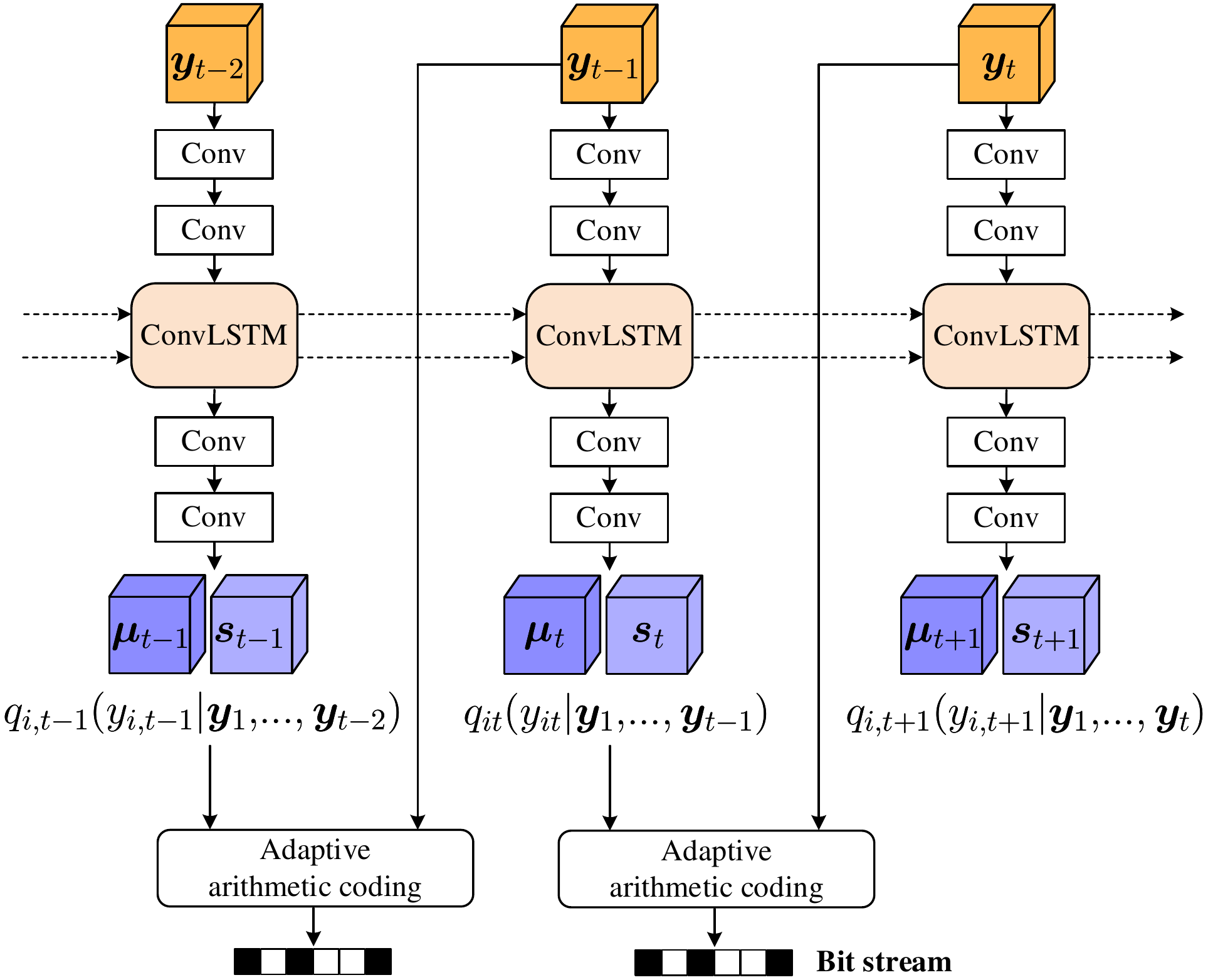}
\caption{The architecture of the RPM network, in which all layers have 128 convolutional filters with the size of $3\times 3$.}\label{fig:RPM}
\end{figure}

Specifically,
adaptive arithmetic coding~\cite{langdon1984introduction} allows to change the PMF for each element in $\bm y_t$, and thus we estimate different conditional PMFs $q_{it}(y_{it}\, | \,\bm y_1, \dots, \bm y_{t-1})$ for different elements $y_{it}$. Here, $y_{it}$ is defined as the element at the $i$-th 3D \emph{location} in $\bm y_t$, and the conditional PMF of $\bm y_t$ can be expressed as
\begin{equation}\label{prod}
q_t(\bm y_t\, | \,\bm y_1, \dots, \bm y_{t-1}) = \prod_{i=1}^N q_{it}(y_{it}\, | \,\bm y_1, \dots, \bm y_{t-1}),
\end{equation}
in which $N$ denotes the number of 3D positions in $\bm y_t$.
As shown in Fig.~\ref{fig:log}, we model $q_{it}(y_{it}\, | \,\bm y_1, \dots, \bm y_{t-1})$ of each element as discretized logistic distribution in our approach. Since the quantization operation in RAE quantizes all $\tilde{y}_{it}\in[y_{it} - 0.5, y_{it} + 0.5)$ to a discrete value $y_{it}$, the conditional PMF of the quantized $y_{it}$ can be obtained by integrating the continuous logistic distribution~\cite{balakrishnan1991handbook} from $(y_{it} - 0.5)$ to $(y_{it} + 0.5)$:
\begin{equation}\label{pmf0}
    q_{it}(y_{it}\, | \, \bm y_1, \dots, \bm y_{t-1})
    = \int_{y_{it}-0.5}^{y_{it}+0.5} \text{Logistic}(y; \mu_{it}, s_{it}) dy,
\end{equation}
in which the logistic distribution is defined as 
\begin{equation}\label{logistic}
\text{Logistic}(y; \mu, s)=\frac{\exp(-(y-\mu)/s)}{s(1+\exp(-(y-\mu)/s))^2}, 
\end{equation} 
and its integral is the sigmoid distribution, \ie,
\begin{equation}\label{sigmoid}
\int\text{Logistic}(y; \mu, s)dy=\text{Sigmod}(y; \mu, s)+C.
\end{equation} 
Given \eqref{pmf0}, \eqref{logistic} and \eqref{sigmoid}, the estimated conditional PMF can be simplified as
\begin{equation}\label{pmf}
\begin{aligned}
   q_{it}(y_{it}\, | \, \bm y_1, \dots, \bm y_{t-1})
    &= \text{Sigmoid}(y_{it} + 0.5; \mu_{it}, s_{it}) \\ & \quad - \text{Sigmoid}(y_{it} - 0.5; \mu_{it}, s_{it}).
\end{aligned}
\end{equation}
It can be seen from \eqref{pmf}, the conditional PMF at each \textit{location} is modelled with parameters $\mu_{it}$ and $s_{it}$, which are varying for different locations in $\bm{y}_t$.
The RPM network is proposed to recurrently estimate $\bm{\mu}_{t} = \{\mu_{it}\}_{i=1}^N$ and $\bm{s}_{t} = \{s_{it}\}_{i=1}^N$ in \eqref{pmf}.
Fig.~\ref{fig:RPM} demonstrates the detailed architecture of our RPM network, which contains a recurrent network $P$ with convolution layers and a ConvLSTM cell in the middle. Due to the recurrent structure, $\bm{\mu}_{t}$ and $\bm{s}_{t}$ are generated based on all previous latent representations, \ie, 
\begin{equation}\label{mus}
    \bm{\mu}_{t}, \bm{s}_{t} = P(\bm y_1, \dots, \bm y_{t-1}; \bm \theta_P),
\end{equation}
where $\bm \theta_P$ represents the trainable parameters in RPM. Because $P$ takes previous latent representations $\bm y_1, \dots, \bm y_{t-1}$ as inputs, $\bm{\mu}_{t}$ and $\bm{s}_{t}$ learn to model the probability of each $y_{it}$ conditioned on $\bm y_1, \dots, \bm y_{t-1}$ according to \eqref{pmf}. Finally, the conditional PMFs $q_{it}(y_{it}\, | \, \bm y_1, \dots, \bm y_{t-1} )$ are applied to the adaptive arithmetic coding~\cite{langdon1984introduction} to encode $\bm{y}_t$ into a bit stream.

\subsection{Training}\label{train}
In this paper, we utilize the Multi-Scale Structural SIMilarity (MS-SSIM) index and the Peak Signal-to-Noise Ratio (PSNR) to evaluate compression quality, and train two models optimized for MS-SSIM and PSNR, respectively. The distortion $D$ is defined as $1 - \text{MS-SSIM}$ when optimizing for MS-SSIM, and as the Mean Square Error (MSE) when training the PSNR model. As Fig.~\ref{fig:framework} shows, our approach uses the uni-directional Low-Delay P (LDP) structure. We follow \cite{yang2020heirarchical} to compress the I-frame $\bm f_0$ with the learned image compression method~\cite{lee2019context} for the MS-SSIM model, and with BPG~\cite{BPG} for the PSNR model. Because of lacking previous latent representation for the first P-frame $\bm f_1$, $\bm y^m_1$ and $\bm y^r_1$ are compressed by the spatial entropy model of \cite{balle2017end}, with the bit-rate defined as $R_1(\bm y^m_1)$ and $R_1(\bm y^r_1)$, respectively. The following P-frames are compressed with the proposed RPM network. For $t\geq2$, the actual bit-rate can be calculated as
\begin{equation}
    \begin{aligned}
    R_{\text{RPM}}(\bm y_t) 
    &= -\log_2(q_t(\bm y_t\, | \,\bm y_1, \dots, \bm y_{t-1})) \\
    &=\sum_{i=1}^N -\log_2(q_{it}(y_{it}\, | \,\bm y_1, \dots, \bm y_{t-1})),
    \end{aligned}
\end{equation}
in which $q_t(\bm y_t\, | \,\bm y_1, \dots, \bm y_{t-1})$ is modelled by the proposed RPM according to \eqref{prod} to \eqref{mus}. Note that, assuming that the distribution of the training set is identical with the true distribution, the actual bit-rate $R_{\text{RPM}}(\bm y_t)$ is expected to be the conditional cross entropy in \eqref{CE_new}. In our approach, two RPM networks are applied to the latent representations of motion and residual, and their bit-rates are defined as $R_{\text{RPM}}(\bm y^m_t)$ and $R_{\text{RPM}}(\bm y^r_t)$, respectively. 

\begin{figure*}[!t]
\centering
\includegraphics[width=.99\linewidth]{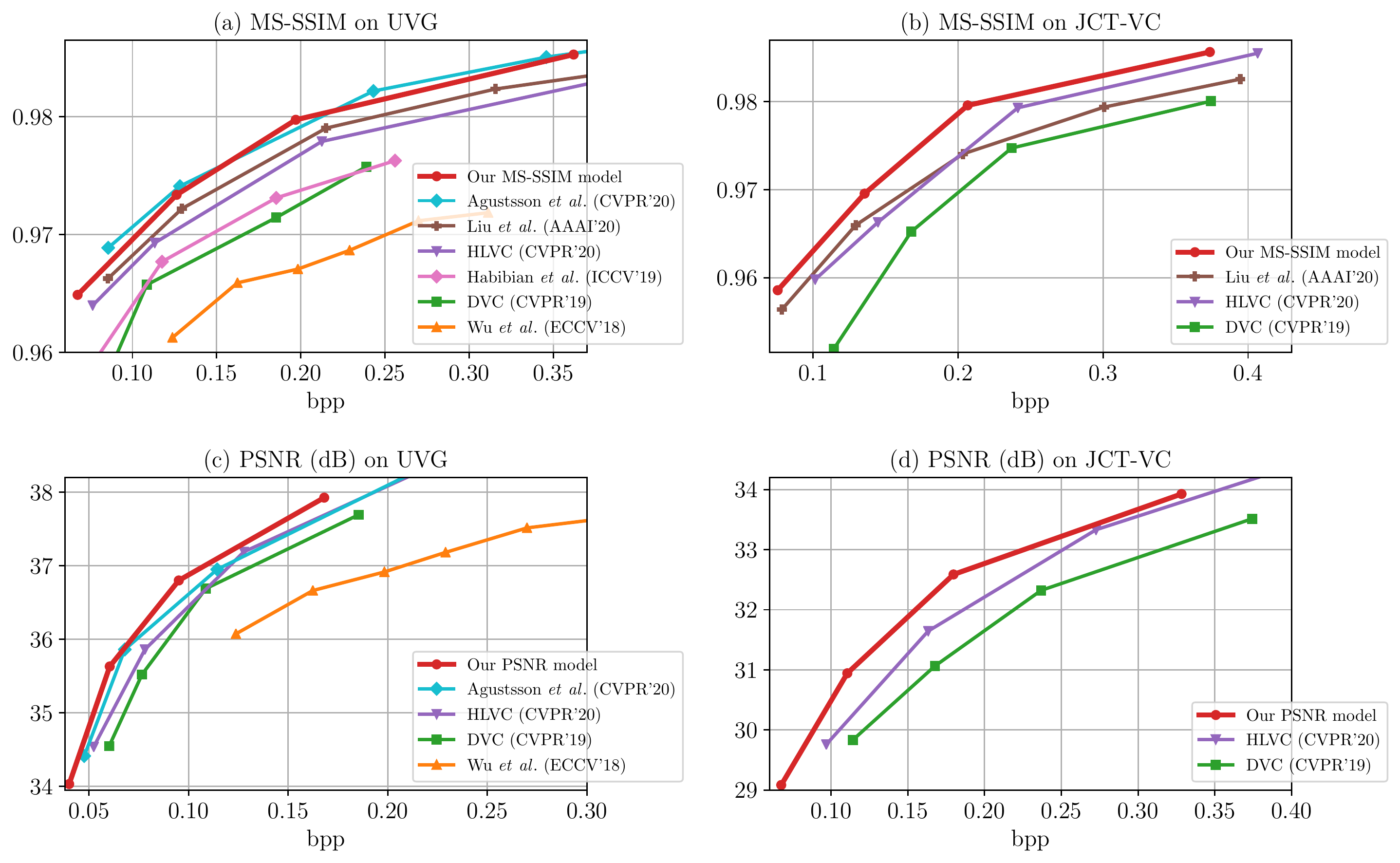}
\caption{The rate-distortion performance of our RLVC approach compared with the learned video compression approaches on the UVG and JCT-VC datasets.} \label{fig:curve}
\end{figure*}
\begin{figure*}[!t]
\centering
\includegraphics[width=.99\linewidth]{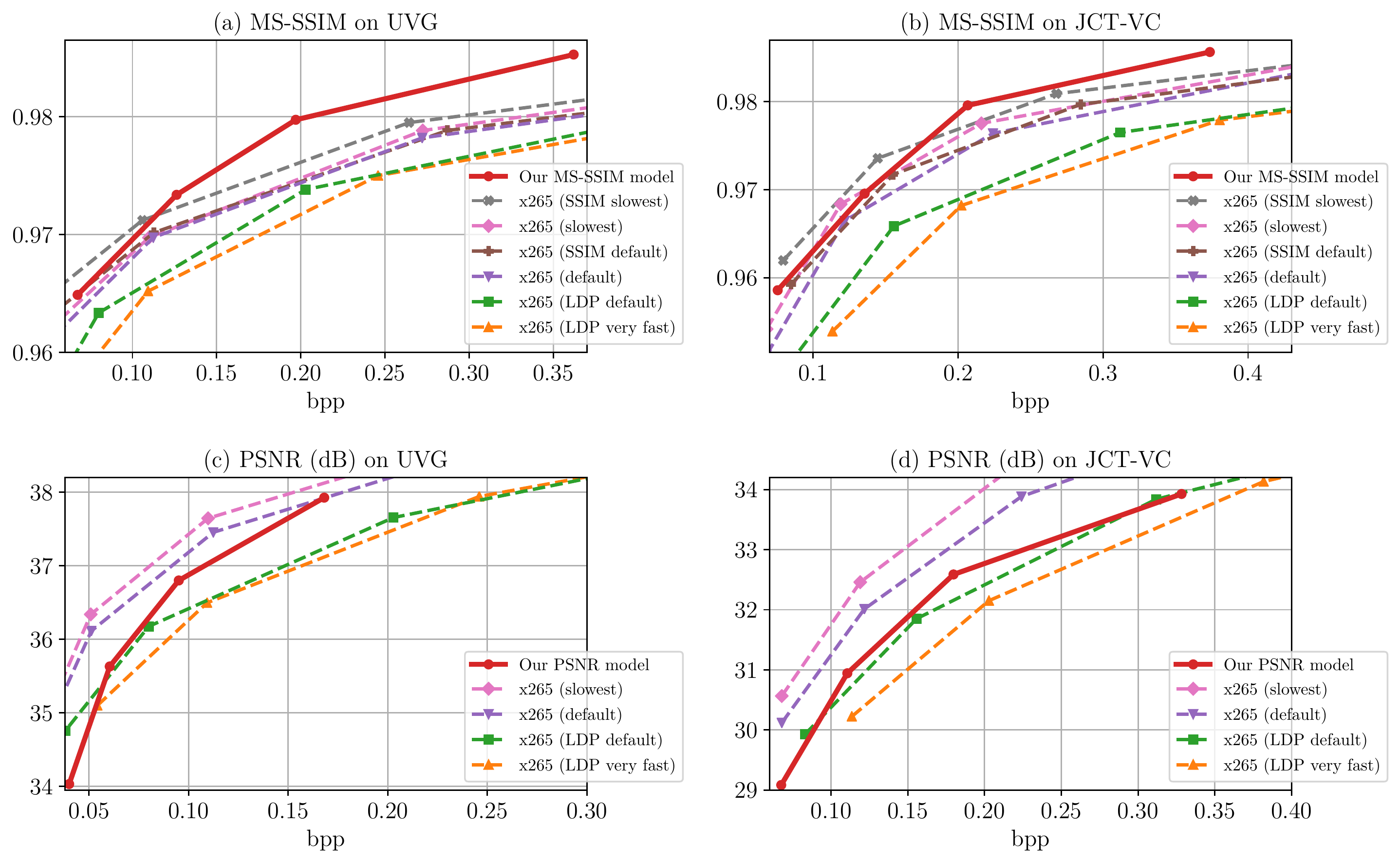}
\caption{The rate-distortion performance of our RLVC approach compared with different settings of x265 on the UVG and JCT-VC datasets.} \label{fig:curve_x265}
\end{figure*}

Our RLVC approach is trained on the Vimeo-90k~\cite{xue2019video} dataset, in which each training sample has 7 frames. The first frame is compressed as the I-frame and the other 6 frames are P-frames. First, we warm up the network on the first P-frame $\bm f_1$ in a progressive manner. At the beginning, the motion estimation network is trained with the loss function of
\begin{equation}\label{loss_me}
    \mathcal{L}_{\text{ME}} = D(\bm f_1, W(\bm f_0, \bm x^m_1)),
\end{equation}
in which $\bm x^m_1$ is the output of the motion estimation network (as shown in Fig.~\ref{fig:framework}) and $W$ is the warping operation. When $L_{\text{ME}}$ is converged, we further include the RAE network for compressing motion and the motion compensation network into training, using the following loss function
\begin{equation}\label{loss_mc}
    \mathcal{L}_{\text{MC}} = \lambda\cdot D(\bm f_1, \bm f'_1) + R_1(\bm y^m_1).
\end{equation}
After the convergence of $L_{\text{MC}}$, the whole network is jointly trained on $\bm f_1$ by the loss of
\begin{equation}\label{loss1}
    \mathcal{L}_{1} = \lambda\cdot D(\bm f_1, \bm{\hat{f}}_1) + R_1(\bm y^m_1) + R_1(\bm y^r_1).
\end{equation}
In the following, we train our recurrent model in an end-to-end manner on the sequential training frames using loss function of
\begin{equation}\label{loss}
\begin{aligned}
    \mathcal{L} &= \lambda\cdot \sum_{t=1}^6 D(\bm f_t, \bm{\hat{f}}_t) \\
    &\quad+ R_1(\bm y^m_t) + R_1(\bm y^r_t) 
    + \sum_{t=2}^6 \Big(R_{\text{RPM}}(\bm y^m_t) + R_{\text{RPM}}(\bm y^r_t)\Big).
\end{aligned}
\end{equation}

During training, quantization is relaxed by the method in \cite{balle2017end} to avoid zero gradients. We follow \cite{yang2020heirarchical} to set $\lambda$ as 8, 16, 32 and 64 for MS-SSIM, and as 256, 512, 1024 and 2048 for PSNR. The Adam optimizer~\cite{kingma2014adam} is utilized for training. The initial learning rate is set as $10^{-4}$ for all loss functions \eqref{loss_me}, \eqref{loss_mc}, \eqref{loss1} and \eqref{loss}. When training the whole model by the final loss of \eqref{loss}, we decrease the learning rate after convergence by the factor of 10 until $10^{-6}$.

\section{Experiments}\label{exp}
\subsection{Settings}\label{setting}

\begin{table*}[!t]
\centering
\caption{BDBR calculated by \textbf{MS-SSIM} with the anchor of x265 (LDP very fast). \textbf{Bold} is the best results in learned approaches.}\label{tab:bdbr_ssim}
\setlength{\tabcolsep}{4.8pt}

\begin{tabular}{clrrrrr@{\hskip 5ex}rrrrr}
\cmidrule[\heavyrulewidth]{1-12}  %
& & \multicolumn{5}{c}{Learned}&\multicolumn{5}{c}{Non-learned} \\
\cmidrule(r{1.5em}){3-7}
\cmidrule(l{.5em}){8-12} 

& & \multicolumn{1}{c}{DVC} &
\multicolumn{1}{c}{Cheng} &
\multicolumn{1}{c}{Habibian}& \multicolumn{1}{c}{HLVC}&
\multicolumn{1}{l}{RLVC} & \multicolumn{1}{c}{x265} & \multicolumn{1}{c}{x265} & 
\multicolumn{1}{c}{x265} &
\multicolumn{1}{c}{x265} &
\multicolumn{1}{c}{x265} \\
Dataset & \multicolumn{1}{c}{Video} & \multicolumn{1}{c}{\cite{lu2019dvc}}&  
\multicolumn{1}{c}{\cite{cheng2019learning}}&  
\multicolumn{1}{c}{\cite{habibian2019video}}&  \multicolumn{1}{c}{\cite{yang2020heirarchical}}&
\multicolumn{1}{l}{(Ours)}&
\multicolumn{1}{c}{LDP def.}&
\multicolumn{1}{c}{default}&
\multicolumn{1}{c}{SSIM def.}&
\multicolumn{1}{c}{slowest}&
\multicolumn{1}{c}{SSIM slowest}
 \\
 
\cmidrule[\heavyrulewidth]{1-12}
\multirow{8}[0]{*}{UVG} & \multicolumn{1}{l}{\textit{Beauty}} & $ -14.85  $ & \multicolumn{1}{c}{-} & $ -44.63  $ & $ -41.39  $ & $ \mathbf{-49.22} $ & $ -3.35  $ & $ 3.18  $ & $ -0.76  $ & $ 6.31  $ & $ -23.72  $\\
          &  \multicolumn{1}{l}{\textit{Bosphorus}} & $ 10.03  $ & \multicolumn{1}{c}{-} & $ -13.77  $ & $ -51.22  $ & $ \mathbf{-62.02}  $ & $ -2.63  $ & $ -45.35  $ & $ -48.07  $ & $ -46.01  $ & $ -55.27  $\\
          &  \multicolumn{1}{l}{\textit{HoneyBee}}  & $ -21.63  $ & \multicolumn{1}{c}{-} & $ -4.13  $ & $ -42.87  $ & $ \mathbf{-43.49}  $ & $ -54.90  $ & $ -70.78  $ & $ -67.57  $ & $ -66.96  $ & $ -66.58  $\\
          &  \multicolumn{1}{l}{\textit{Jockey}}  & $ 104.82  $ & \multicolumn{1}{c}{-} & $ 56.38  $ & $ 6.97  $ & $ \mathbf{-12.54}  $ & $ -13.41  $ & $ -15.15  $ & $ -27.32  $ & $ -20.98  $ & $ -44.95  $\\
          &  \multicolumn{1}{l}{\textit{ReadySetGo}}  & $ 2.77  $ & \multicolumn{1}{c}{-} & $ 89.06  $ & $ -7.32  $ & $ \mathbf{-20.98} $ & $ -13.54  $ & $ -36.94  $ & $ -40.96  $ & $ -43.07  $ & $ -52.11  $\\
          &  \multicolumn{1}{l}{\textit{ShakeNDry}}  & $ -20.94  $ & \multicolumn{1}{c}{-} & $ -35.10  $ & $ -32.82  $ & $ \mathbf{-40.10}  $ & $ -24.08  $ & $ -38.64  $ & $ -40.96  $ & $ -45.02  $ & $ -51.36  $\\
          &  \multicolumn{1}{l}{\textit{YachtRide}}  & $ -3.83  $ & \multicolumn{1}{c}{-} & $ -21.85  $ & $ -42.17  $ & $ \mathbf{-55.96}  $ & $ -0.09  $ & $ -20.76  $ & $ -23.32  $ & $ -25.68  $ & $ -31.69  $\\
    \cmidrule{2-12}
          & \multicolumn{1}{l}{\textbf{Average}} & $ 8.05  $ & \multicolumn{1}{c}{-} & $ 3.71  $ & $ -30.12  $ & $ \mathbf{-40.62}  $ & $ -16.00  $ & $ -32.06  $ & $ -35.57  $ & $ -34.49  $ & $ -46.52  $\\
    \cmidrule[\heavyrulewidth]{1-12}
    \multicolumn{1}{c}{\multirowcell{6}{JCT-VC \\ Class B}} &  \multicolumn{1}{l}{\textit{BasketballDrive}}  & $ 15.47  $ & \multicolumn{1}{c}{-} &\multicolumn{1}{c}{-}& $ -34.98  $ & $ \mathbf{-48.10}  $ & $ 2.19  $ & $ -18.05  $ & $ -30.26  $ & $ -22.88  $ & $ -42.68  $\\
          &  \multicolumn{1}{l}{\textit{BQTerrace}}  & $ 15.08  $ & \multicolumn{1}{c}{-} & \multicolumn{1}{c}{-} & $ -22.52  $ & $ \mathbf{-44.10}  $ & $ -30.97  $ & $ -55.70  $ & $ -50.36  $ & $ -56.55  $ & $ -57.09  $\\
          &  \multicolumn{1}{l}{\textit{Cactus}}  & $ -21.40  $ & \multicolumn{1}{c}{-} & \multicolumn{1}{c}{-} & $ -43.63  $ & $ \mathbf{-53.96}  $ & $ -26.22  $ & $ -41.15  $ & $ -45.28  $ & $ -45.32  $ & $ -52.98  $\\
          &  \multicolumn{1}{l}{\textit{Kimono}}  & $ -2.67  $ & \multicolumn{1}{c}{-}& \multicolumn{1}{c}{-}& $ -46.79  $ & $ \mathbf{-56.73}  $ & $ -7.24  $ & $ -13.57  $ & $ -25.03  $ & $ -18.63  $ & $ -34.77  $\\
          &  \multicolumn{1}{l}{\textit{ParkScene}}  & $ -20.17  $ & \multicolumn{1}{c}{-} & \multicolumn{1}{c}{-}& $ -39.31  $ & $ \mathbf{-49.18}  $ & $ -9.46  $ & $ -43.61  $ & $ -47.04  $ & $ -48.54  $ & $ -55.94  $\\
          \cmidrule{2-12}
          &  \multicolumn{1}{l}{\textbf{Average}}  & $ -2.74  $ & \multicolumn{1}{c}{-} &\multicolumn{1}{c}{-} & $ -37.44  $ & $ \mathbf{-50.42}  $ & $ -14.34  $ & $ -34.42  $ & $ -39.60  $ & $ -38.38  $ & $ -48.69  $\\
           \cmidrule[\heavyrulewidth]{1-12}
    \multicolumn{1}{c}{\multirowcell{5}{JCT-VC \\  Class C}} &  \multicolumn{1}{l}{\textit{BasketballDrill}}  & $ 5.54  $ & $ 17.97  $ & \multicolumn{1}{c}{-} & $ -18.45  $ & $ \mathbf{-32.57}  $ & $ -18.59  $ & $ -41.12  $ & $ -42.70  $ & $ -46.53  $ & $ -50.39  $\\
          &  \multicolumn{1}{l}{\textit{BQMall}}  & $ 4.84  $ & $ \mathbf{-38.59}  $ & \multicolumn{1}{c}{-} & $ -20.33  $ & $ -36.88  $ & $ -12.56  $ & $ -34.05  $ & $ -39.04  $ & $ -43.13  $ & $ -50.79  $\\
          &  \multicolumn{1}{l}{\textit{PartyScene}}  & $ -23.60  $ & $ -6.53  $ & \multicolumn{1}{c}{-}& $ -30.29  $ & $ \mathbf{-38.81}  $ & $ -11.70  $ & $ -41.53  $ & $ -42.67  $ & $ -48.07  $ & $ -52.30  $\\
          & \multicolumn{1}{l}{\textit{RaceHorses} (480p)}  & $ -14.29  $ & $ 41.07  $ & \multicolumn{1}{c}{-} & $ -25.45  $ & $ \mathbf{-35.50}  $ & $ -8.08  $ & $ -21.69  $ & $ -22.89  $ & $ -30.82  $ & $ -36.61  $\\
          \cmidrule{2-12}
          &  \multicolumn{1}{l}{\textbf{Average}}  & $ -6.88  $ & $ 3.48  $ & \multicolumn{1}{c}{-} & $ -23.63  $ & $ \mathbf{-35.94}  $ & $ -12.73  $ & $ -34.60  $ & $ -36.82  $ & $ -42.14  $ & $ -47.52  $\\
           \cmidrule[\heavyrulewidth]{1-12}
    \multicolumn{1}{c}{\multirowcell{5}{JCT-VC \\  Class D}} & \multicolumn{1}{l}{\textit{BasketballPass}}  & $ 0.67  $ & $ -44.96  $ & \multicolumn{1}{c}{-} & $ -36.24  $ & $ \mathbf{-51.40}  $ & $ -14.45  $ & $ -32.55  $ & $ -32.89  $ & $ -39.03  $ & $ -42.69  $\\
          & \multicolumn{1}{l}{\textit{BlowingBubbles}} & $ -29.38  $ & $ -22.92  $ & \multicolumn{1}{c}{-} & $ -39.84  $ & $ \mathbf{-49.57}  $ & $ -11.04  $ & $ -39.02  $ & $ -41.12  $ & $ -45.49  $ & $ -51.35  $\\
          & \multicolumn{1}{l}{\textit{BQSquare}} & $ -25.50  $ & $ -39.60  $ & \multicolumn{1}{c}{-} & $ \mathbf{-97.56}  $ & $ -44.71  $ & $ -14.16  $ & $ -57.31  $ & $ -56.04  $ & $ -62.24  $ & $ -60.37  $\\
          & \multicolumn{1}{l}{\textit{RaceHorses} (240p)} & $ -19.82  $ & $ 12.60  $ & \multicolumn{1}{c}{-} & $ -36.59  $ & $ \mathbf{-49.75}  $ & $ -7.63  $ & $ -23.15  $ & $ -24.70  $ & $ -37.22  $ & $ -41.15  $\\
          \cmidrule{2-12}
          & \multicolumn{1}{l}{\textbf{Average}} & $ -18.51  $ & $ -23.72  $ &\multicolumn{1}{c}{-} & $ \mathbf{-52.56}  $ & $ -48.85  $ & $ -11.82  $ & $ -38.01  $ & $ -38.69  $ & $ -45.99  $ & $ -48.89  $\\
           \cmidrule[\heavyrulewidth]{1-12}
    \multicolumn{2}{c}{\textbf{Average on all videos}} & $ -2.94  $ & \multicolumn{1}{c}{-} & \multicolumn{1}{c}{-} & $ -35.14  $ & $ \mathbf{-43.78}  $ & $ -14.10  $ & $ -34.35  $ & $ -37.45  $ & $ -39.29  $ & $ -47.74  $\\
     \cmidrule[\heavyrulewidth]{1-12}

\end{tabular}
\end{table*}

The experiments are conducted to validate the effectiveness of our RLVC approach. We evaluate the performance on the same test set as \cite{yang2020heirarchical}, \ie, the JCT-VC~\cite{bossen2013common} (Classes B, C and D) and the UVG~\cite{UVG} datasets.
The JCT-VC Class B and UVG are high resolution ($1920\times 1080$) datasets, and the JCT-VC Classes C and D are with the resolution of $832\times 480$ and $416\times 240$, respectively.
We compare our RLVC approach with the latest learned video compression methods: HLVC~\cite{yang2020heirarchical} (CVPR'20), Liu~\etal~\cite{liu2019learned} (AAAI'20), Habibian~\etal~\cite{habibian2019video} (ICCV'19), DVC~\cite{lu2019dvc} (CVPR'19), Cheng~\etal~\cite{cheng2019learning} (CVPR'19) and Wu~\etal~\cite{wu2018video} (ECCV'18).
To compare with the handcrafted video coding standard H.265~\cite{sullivan2012overview}, we first include the \textit{LDP very fast} setting of x265 into comparison, which is used as the anchor in previous learned compression works \cite{lu2019dvc, yang2020heirarchical, liu2019learned}. We also compare our approach with the \textit{LDP default}, the \textit{default} and the \textit{slowest} settings of x265. Moreover, the \textit{SSIM-tuned} x265 is also compared with our MS-SSIM model. The detailed configurations of x265 are listed as follows:

\begin{flushleft}
\begin{itemize}
    \item \textbf{x265 (LDP very fast):} 
    
    \texttt{ffmpeg (input) -c:v libx265\\
    -preset veryfast -tune zerolatency \\
    -x265-params "crf=CRF:keyint=10" output.mkv}
    
    \vspace{.5em}
    \item \textbf{x265 (LDP default): }
    
    \texttt{ffmpeg (input) -c:v libx265 \\
    -tune zerolatency \\
    -x265-params "crf=CRF" output.mkv}
    
    \vspace{.5em}
    \item \textbf{x265 (default):} 
    
    \texttt{ffmpeg (input) -c:v libx265 \\
    -x265-params "crf=CRF" output.mkv}
    
    \vspace{.5em}
    \item \textbf{x265 (SSIM default):} 
    
    \texttt{ffmpeg (input) -c:v libx265
    -tune ssim \\
    -x265-params "crf=CRF" output.mkv}
    
    \vspace{.5em}
    \item \textbf{x265 (slowest):} 
    
    \texttt{ffmpeg (input) -c:v libx265 \\ -preset placebo\footnote{ \textit{Placebo} is the slowest setting among the 10 speed levels in x265.}\\
    -x265-params "crf=CRF" output.mkv}
    
    \vspace{.5em}
    \item \textbf{x265 (SSIM slowest):} 
    
    \texttt{ffmpeg (input) -c:v libx265 \\
    -preset placebo -tune ssim\\
    -x265-params "crf=CRF" output.mkv}

\end{itemize}
\end{flushleft}
In above settings, ``\texttt{(input)}'' is short for ``\texttt{-pix\_fmt yuv420p -s WidthxHeight -r Framerate  -i  input.yuv}''. CRF indicates the compression quality, and lower CRF corresponds to higher quality. We set CRF = 15, 19, 23, 27 for the JCT-VC dataset, and set CRF = 7, 11, 15, 19, 23 for the UVG dataset. 

Please refer to the \textit{Supporting Document} for the experimental results on more datasets, such as the conversational video dataset and the MCL-JCV~\cite{wang2016mcl} dataset.

\subsection{Performance}\label{perf}

\textbf{Comparison with learned approaches.} Fig.~\ref{fig:curve} illustrates the rate-distortion curves of our RLVC approach in comparison with previous learned video compression approaches on the UVG and JCT-VC datasets. Among the compared approaches, Liu~\etal~\cite{liu2019learned} and Habibian~\etal~\cite{habibian2019video} are optimized for MS-SSIM. DVC~\cite{lu2019dvc} and Wu~\etal~\cite{wu2018video} are optimized for PSNR. HLVC~\cite{yang2020heirarchical} and Agustsson~\etal~\cite{agustsson2020scale} train two models for MS-SSIM and PSNR, respectively. 
As we can see from Fig.~\ref{fig:curve} (a) and (b), our MS-SSIM model performs competitively to Agustsson~\etal~\cite{agustsson2020scale}, and outperforms all other learned approaches, including the state-of-the-art MS-SSIM optimized approaches Liu~\etal~\cite{liu2019learned} (AAAI'20), HLVC~\cite{yang2020heirarchical} (CVPR'20) and Habibian~\etal~\cite{habibian2019video} (ICCV'19). In terms of PSNR, Fig.~\ref{fig:curve} (c) and (d) indicate the superior performance of our PSNR model to the PSNR optimized models Agustsson~\etal~\cite{agustsson2020scale}, HLVC~\cite{yang2020heirarchical} (CVPR'20), DVC~\cite{lu2019dvc} (CVPR'19) and Wu~\etal~\cite{wu2018video} (ECCV'18). It is worth pointing out that our RLVC approach employs the same motion estimation network as HLVC~\cite{yang2020heirarchical} and DVC~\cite{lu2019dvc}, and applying the space-scale motion proposed in \cite{agustsson2020scale} may further improve the performance of RLVC.

We further tabulate the Bj{\o}ntegaard Delta Bit-Rate (BDBR)~\cite{bjontegaard} results calculated by MS-SSIM and PSNR with the anchor of x265 (LDP very fast) in Tables~\ref{tab:bdbr_ssim} and \ref{tab:bdbr_psnr}, respectively.\footnote{Since \cite{wu2018video,liu2019learned} do not release the results on each video, their BDBR values cannot be obtained.} Note that, BDBR calculates the average bit-rate difference in comparison with the anchor. Lower BDBR value indicates better performance, and negative BDBR indicates saving bit-rate in comparison with the anchor, \ie, outperforming the anchor. In Tables~\ref{tab:bdbr_ssim} and \ref{tab:bdbr_psnr}, the bold numbers are the best results in learned approaches. As Table~\ref{tab:bdbr_ssim} shows, in terms of MS-SSIM, the proposed RLVC approach outperforms previous learned approaches on all videos in the high resolution datasets UVG and JCT-VC Class B. In all the 20 test videos, we achieve the best results in learned approaches on 18 videos, and have the best average BDBR performance among all learned approaches. Moreover, Table~\ref{tab:bdbr_psnr} shows that, in terms of PSNR, our PSNR model has better performance than all existing learned approaches on all test videos.  

Note that, the latest HLVC~\cite{yang2020heirarchical} (CVPR'20) approach introduces bi-directional prediction, hierarchical structure and post-processing into learned video compression, while the proposed RLVC approach only works in the uni-directional IPPP model without post-processing (as shown in Fig.~\ref{fig:framework}). Nevertheless, our approach still achieves better performance than HLVC~\cite{yang2020heirarchical}, validating the effectiveness of our recurrent compression framework with the proposed RAE and RPM networks.

\begin{table}[!t]
\centering
\caption{BDBR calculated by \textbf{PSNR} with the anchor of x265 (LDP very fast). \textbf{Bold} is the best results in learned approaches.}\label{tab:bdbr_psnr}
\setlength{\tabcolsep}{2pt}

\begin{tabular}{p{1.95cm}rrr@{\hskip 2ex}rrr}
\cmidrule[\heavyrulewidth]{1-7}  %
& \multicolumn{3}{c}{Learned}&\multicolumn{3}{c}{Non-learned} \\
\cmidrule(r{.5em}){2-4}
\cmidrule(l{.5em}){5-7} 

& \multicolumn{1}{c}{DVC} &
\multicolumn{1}{c}{HLVC}&
\multicolumn{1}{l}{RLVC} & \multicolumn{1}{c}{x265} & 
\multicolumn{1}{c}{x265} &
\multicolumn{1}{c}{x265} \\
\multicolumn{1}{c}{Video} & \multicolumn{1}{c}{\cite{lu2019dvc}}&  
 \multicolumn{1}{c}{\cite{yang2020heirarchical}}&
\multicolumn{1}{l}{(Ours)}&
\multicolumn{1}{c}{LDP def.}&
\multicolumn{1}{c}{default}&
\multicolumn{1}{c}{slowest}\\

\cmidrule[\heavyrulewidth]{1-7}

\textit{Beauty} & $ -39.63  $ & $ -48.48  $ & $ \mathbf{-56.46}  $ & $ 3.84  $ & $ 4.01  $ & $ -2.41  $ \\
    \textit{Bosphorus}  & $ 17.57  $ & $ -23.16  $ & $ \mathbf{-35.75} $ & $ -4.06  $ & $ -44.24  $ & $ -47.72  $ \\
    \textit{HoneyBee}  & $ 24.53  $ & $ -26.63  $ & $ \mathbf{-21.98}  $ & $ -48.55  $ & $ -79.03  $ & $ -80.69  $ \\
    \textit{Jockey}  & $ 90.02  $ & $ 105.21  $ & $ \mathbf{82.58}  $ & $ -9.62  $ & $ -21.29  $ & $ -28.96  $ \\
    \textit{ReadySetGo}  & $ 9.03  $ & $ 26.69  $ & $ \mathbf{0.03} $ & $ -12.68  $ & $ -39.76  $ & $ -47.52  $ \\
    \textit{ShakeNDry}  & $ -25.07  $ & $ -26.88  $ & $ \mathbf{-31.52}  $ & $ -21.58  $ & $ -43.43  $ & $ -50.68  $ \\
    \textit{YachtRide}  & $ -14.19  $ & $ -16.34  $ & $ \mathbf{-31.30}  $ & $ -1.95  $ & $ -19.47  $ & $ -27.04  $ \\
\cmidrule{1-7} 
    \textbf{Ave. (UVG)}  & $ 8.89  $ & $ -1.37  $ & $ \mathbf{-13.48}  $ & $ -13.51  $ & $ -34.74  $ & $ -40.72  $ \\
\cmidrule[\heavyrulewidth]{1-7}
    \textit{BasketballDrive}  & $ 35.24  $ & $ 13.21  $ & $ \mathbf{4.40}  $ & $ -1.92  $ & $ -20.70  $ & $ -28.08  $ \\
    \textit{BQTerrace}  & $ 2.28  $ & $ -4.56  $ & $ \mathbf{-20.12}  $ & $ -28.03  $ & $ -60.29  $ & $ -63.44  $ \\
    \textit{Cactus}  & $ -5.19  $ & $ -29.09  $ & $ \mathbf{-34.71}  $ & $ -23.66  $ & $ -48.60  $ & $ -53.60  $ \\
    \textit{Kimono}  & $ -10.79  $ & $ -18.71  $ & $ \mathbf{-34.40}  $ & $ -5.13  $ & $ -15.41  $ & $ -22.46  $ \\
    \textit{ParkScene}  & $ -11.63  $ & $ -19.59  $ & $ \mathbf{-36.16}  $ & $ -7.73  $ & $ -45.64  $ & $ -51.89  $ \\
\cmidrule{1-7} 
    \textbf{Ave. (Class B)}  & $ 1.98  $ & $ -11.75  $ & $ \mathbf{-24.20}  $ & $ -13.29  $ & $ -38.13  $ & $ -43.89  $ \\
\cmidrule[\heavyrulewidth]{1-7}
    \textit{BasketballDrill}  & $ 18.03  $ & $ -3.67  $ & $ \mathbf{-11.75}  $ & $ -21.41  $ & $ -42.21  $ & $ -50.16  $ \\
    \textit{BQMall}  & $ 62.28  $ & $ 13.68  $ & $ \mathbf{-0.32}  $ & $ -12.82  $ & $ -35.31  $ & $ -45.86  $ \\
    \textit{PartyScene}  & $ 8.61  $ & $ 2.08  $ & $ \mathbf{-18.03}  $ & $ -9.81  $ & $ -42.35  $ & $ -50.74  $ \\
    \textit{RaceHorses}  & $ 14.61  $ & $ 19.25  $ & $ \mathbf{11.43} $ & $ -8.05  $ & $ -20.53  $ & $ -30.83  $ \\
\cmidrule{1-7} 
    \textbf{Ave. (Class C)}  & $ 25.88  $ & $ 7.83  $ & $ \mathbf{-4.67}  $ & $ -13.02  $ & $ -35.10  $ & $ -44.40  $ \\
\cmidrule[\heavyrulewidth]{1-7}
    \textit{BasketballPass}  & $ 42.34  $ & $ -3.44  $ & $ \mathbf{-19.16}  $ & $ -17.16  $ & $ -28.73  $ & $ -37.97  $ \\
    \textit{BlowingBubbles} & $ -12.15  $ & $ -19.19  $ & $ \mathbf{-31.67}  $ & $ -10.96  $ & $ -38.53  $ & $ -46.52  $ \\
    \textit{BQSquare} & $ 22.01  $ & $ -19.10  $ & $ \mathbf{-35.27}  $ & $ -16.59  $ & $ -58.64  $ & $ -68.40  $ \\
    \textit{RaceHorses} & $ 9.18  $ & $ -8.55  $ & $ \mathbf{-21.93}  $ & $ -7.90  $ & $ -22.43  $ & $ -37.74  $ \\
\cmidrule{1-7} 
    \textbf{Ave. (Class D)} & $ 15.34  $ & $ -12.57  $ & $ \mathbf{-27.01}  $ & $ -13.15  $ & $ -37.08  $ & $ -47.66 $ \\
\cmidrule[\heavyrulewidth]{1-7}
    \textbf{Ave. (all videos)} & $11.85$  & $-4.36$  & $\mathbf{-17.10}$  & $-13.29$  & $-36.13$  & $-43.64$ \\
\cmidrule[\heavyrulewidth]{1-7} 

\end{tabular}
\end{table}

\textbf{Comparison with x265.} The rate-distortion curves compared with different settings of x265 are demonstrated in Fig.~\ref{fig:curve_x265}. As Fig.~\ref{fig:curve_x265} (a) and (b) show, the proposed MS-SSIM model outperforms x265 (LDP very fast), x265 (LDP default), x265 (default) and x265 (SSIM default) on both the UVG and JCT-VC datasets from low to high bit-rates. Besides, in comparison with the slowest setting of x265, we also achieve better performance on UVG and at high bit-rates on JCT-VC. Moreover, at high bit-rates, we even have higher MS-SSIM performance than the SSIM-tuned slowest setting of x265, which can be seen as the best (MS-)SSIM performance that x265 is able to reach.

\begin{figure*}[!t]
\centering
\includegraphics[width=.95\linewidth]{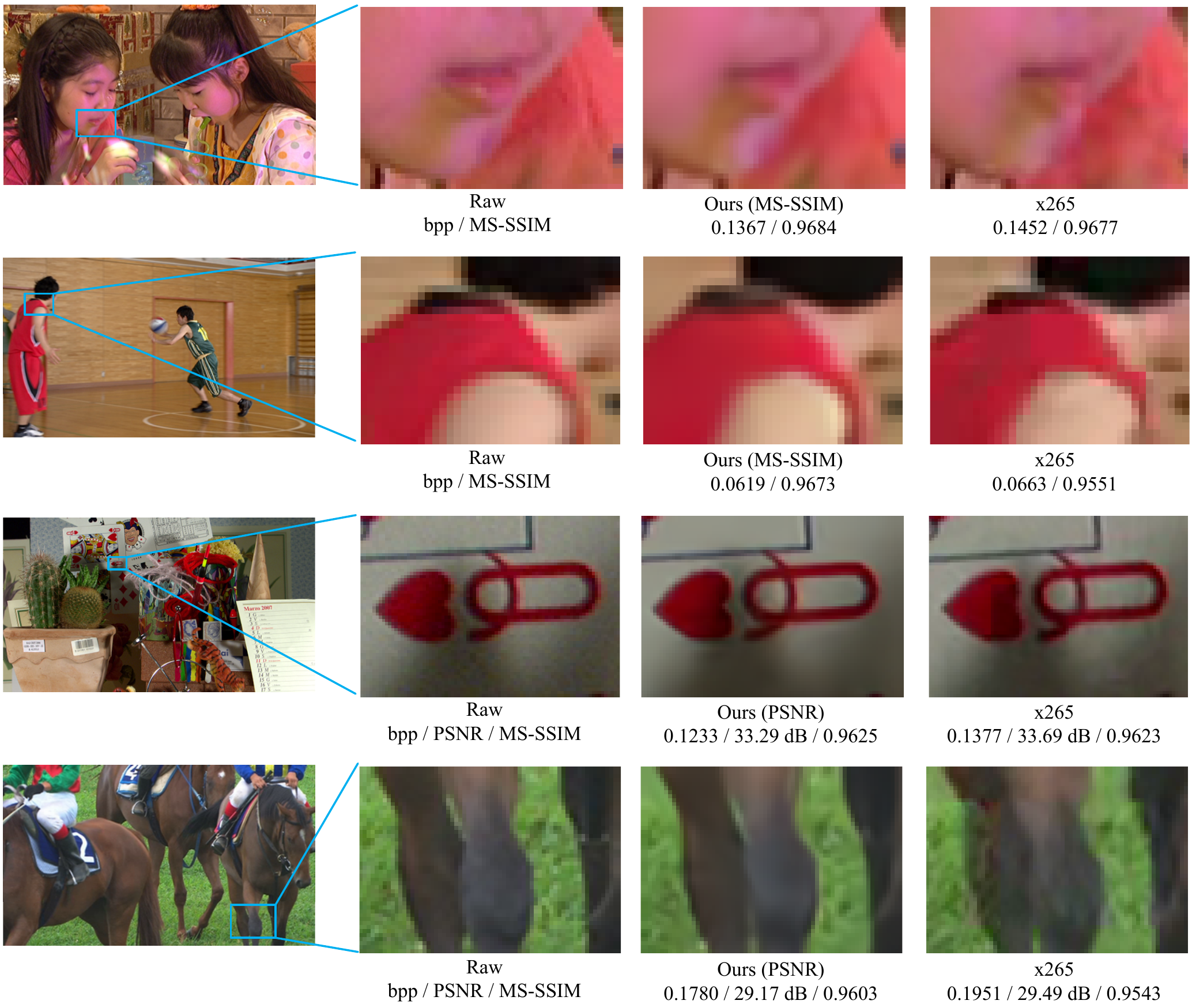}
\caption{The visual results of the MS-SSIM and PSNR models of the proposed RLVC approach in comparison with the default setting of x265.} \label{fig:visual}
\end{figure*}

Similar conclusion can be obtained from the BDBR results calculated by MS-SSIM in Table~\ref{tab:bdbr_ssim}. That is, our RLVC approach averagely reduces $43.78\%$ bit-rate of the anchor x265 (LDP very fast), and outperform x265 (LDP default), x265 (default), x265 (SSIM default) and x265 (slowest). In comparison with x265 (SSIM slowest), we achieve better performance on 8 out of the 20 test videos. We also have better average BDBR result than x265 (SSIM slowest) on JCT-VC Class B, and reach almost the same average performance as x265 (SSIM slowest) on JCT-VC Class D.

In terms of PSNR, Fig.~\ref{fig:curve_x265} (c) and (d) show that our PSNR model outperforms x265 (LDP very fast) from low to high bit-rates on both the UVG and JCT-VC test sets. Besides, we are superior to x265 (LDP default) at high bit-rates on UVG and in a large of bit-rates on JCT-VC. The BDBR results calculated by PSNR in Table~\ref{tab:bdbr_psnr} also indicate that our approach achieves $17.10\%$ less bit-rate than x265 (LDP very fast), and reduces $3.81\%$ more bit-rate than x265 (LDP default). We do not outperform the default and the slowest settings of x265 on PSNR. However, x265 (default) and x265 (slowest) apply advanced strategies in video compression, such as bi-directional prediction and hierarchical frame structure, while our approach only utilizes the IPPP mode. Note that, as far as we know, there is no learned video compression approach beats the default setting of x265 in terms of PSNR. The proposed RLVC approach advances the state-of-the-art learned video compression performance and contributes to catching up with the handcrafted standards step by step.

\begin{table}[!t]
\centering
\caption{Complexity (fps) on 240p videos \\ (including entropy models but not including entropy coding).}\label{tab:complexity}
\begin{tabular}{lcccc}
\cmidrule[\heavyrulewidth]{1-5} 
& \multicolumn{1}{c}{DVC} & \multicolumn{1}{c}{HLVC}&
\multicolumn{1}{l}{Habibian} & \multicolumn{1}{c}{RLVC} \\
 &  \multicolumn{1}{c}{\cite{lu2019dvc}}&   \multicolumn{1}{c}{\cite{yang2020heirarchical}}&
\multicolumn{1}{c}{\cite{habibian2019video}}&
\multicolumn{1}{c}{(Ours)} \\
 \cmidrule{1-5}
 Encoding & 23.3 & 28.8 & 31.3 & 15.9  \\ 
 Decoding & 39.5 & 18.3 & 0.004 & 32.1 \\ 
\cmidrule[\heavyrulewidth]{1-5}
\end{tabular}
\end{table}

\textbf{Visual results.} The visual results of our MS-SSIM and PSNR models are illustrated in Fig.~\ref{fig:visual}, comparing with the default setting of x265. It can be seen from Fig.~\ref{fig:visual} that our MS-SSIM model reaches higher MS-SSIM with lower bit-rate than x265, and produces the compressed frame with less blocky artifacts. For our PSNR model, as discussed above, we do not beat the default setting of x265 in terms of PSNR. However, as Fig.~\ref{fig:visual} shows, our PSNR model also achieves less blocky artifacts and less noise than x265, and is able to reach similar or even higher MS-SSIM than the default setting of x265 in some cases.

\textbf{Computational complexity.} We measure the complexity of the learned approaches on one NVIDIA 1080Ti GPU. The results in terms of frame per second (fps) are shown in Table~\ref{tab:complexity}. 
Note that the data in Table~\ref{tab:complexity} do not include the running time of entropy coding (but include the time for running entropy models), because the official test code\footnote{https://github.com/GuoLusjtu/DVC.} of DVC~\cite{lu2019dvc} does not have the entropy coding module, and the running time of~\cite{habibian2019video} provided by the authors also does not include entropy coding.
As Table~\ref{tab:complexity} shows, due to the recurrent cells in our auto-encoders and probability model, the superior performance of our approach is at the cost of the higher encoding complexity than previous approaches. Nevertheless, we have faster decoding than \cite{yang2020heirarchical,habibian2019video}, and achieve frame rate $> 30$ on 240p videos. Note that, HLVC~\cite{yang2020heirarchical} adopts an enhancement network in the decoder to improve compression quality, which increases decoding complexity. Our RLVC approach (without enhancement) still reaches higher compression performance than HLVC~\cite{yang2020heirarchical}, and also has faster decoding speed. Besides, the auto-regressive (PixelCNN-like) probability model used in \cite{habibian2019video} leads to slow decoding, while the proposed RPM network is more efficient.

\begin{figure}[!t]
\includegraphics[width=.95\linewidth]{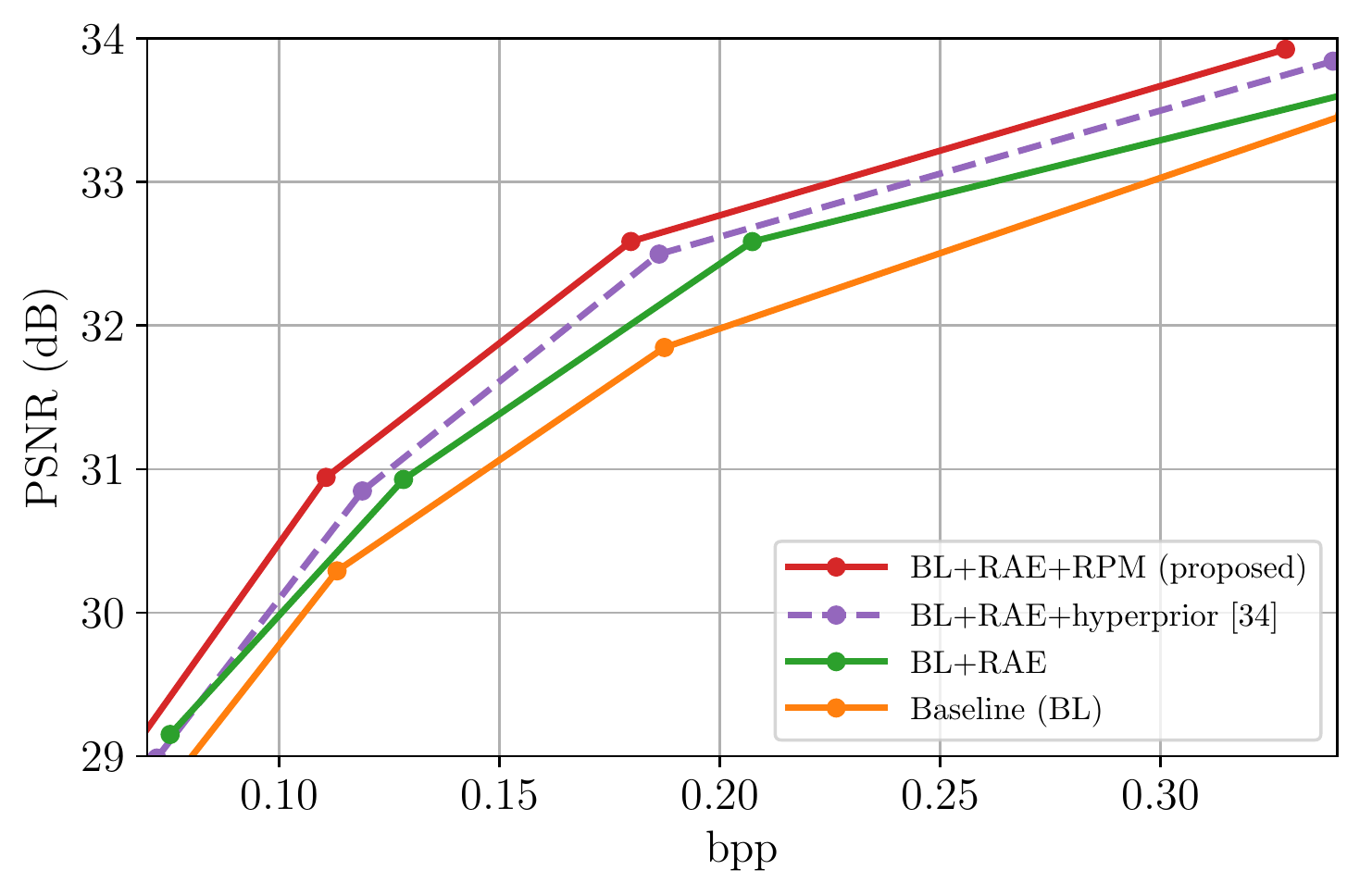}
\caption{Ablation results on the JCT-VC dataset.}\label{fig:abl}
\end{figure}

\subsection{Ablation studies}\label{abl}

The ablation studies are conducted to verify the effectiveness of each recurrent component in our approach. We define the baseline (BL) as our framework without recurrent cells, \ie, without recurrent cells in auto-encoders and replacing our RPM network with the factorized spatial entropy model \cite{balle2017end}. 
In the following, we first enable the recurrent cell in autoencoders, \ie, the proposed RAE network (BL+RAE).
Then, our RPM network is further applied to replace the spatial model \cite{balle2017end} (BL+RAE+RPM, \ie, our full model). Besides, we also compare our RPM network with the hyperprior spatial entropy model~\cite{balle2018variational}.

\textbf{The proposed RAE.} As Fig.~\ref{fig:abl} shows, the rate-distortion curve of BL+RAE outperforms the baseline model. It verifies the effectiveness of the recurrent architecture in the proposed RAE.
It is probably because the dual recurrent cells in both the encoder and decoder learn to encode the residual information between the current and previous inputs, which reduces the information content represented by each latent representation, and then the decoder reconstructs the output based on the encoded residual and previous outputs. This results in efficient compression.

\begin{figure}[!t]
\centering
\subfigure[The proposed RLVC with sequential prior]{\includegraphics[width=1\linewidth]{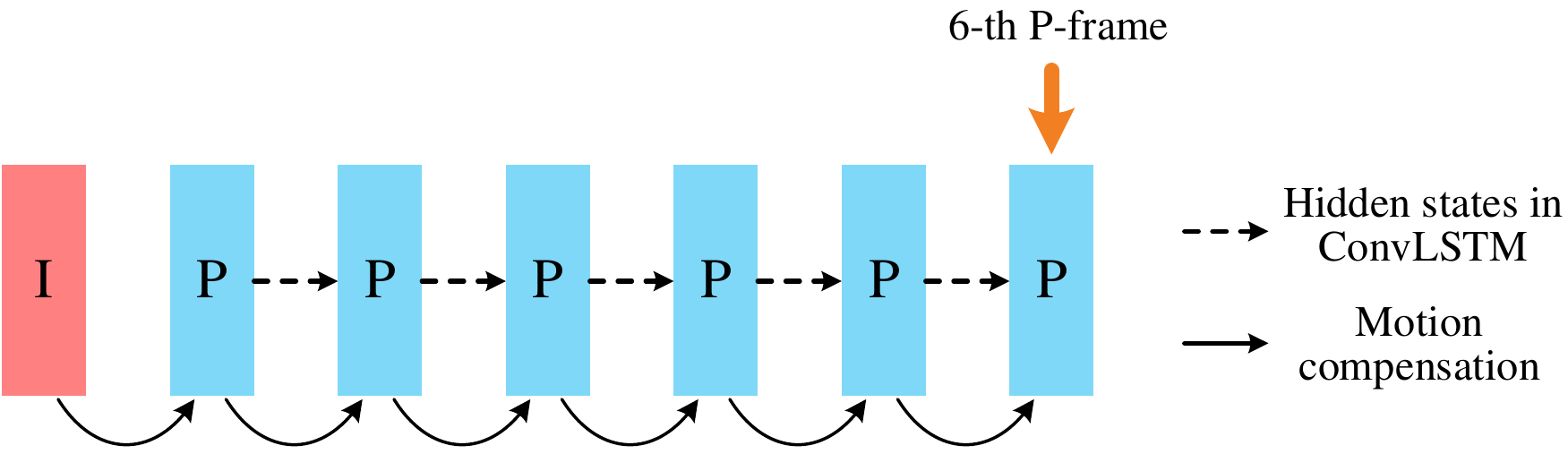}}
\subfigure[Ablation study of one-frame prior]{\includegraphics[width=1\linewidth]{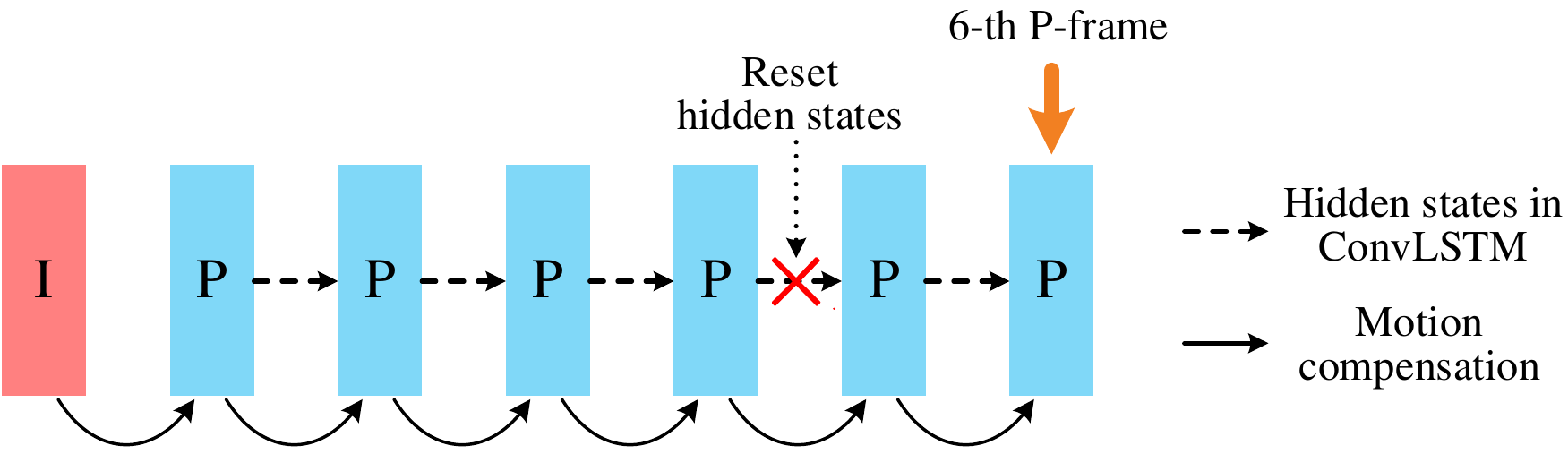}}
\caption{The ablation study of (a) the proposed sequential prior and (b) one-frame prior.}\label{fig:prior}
\end{figure}

\begin{figure}[!t]
\centering
\includegraphics[width=.95\linewidth]{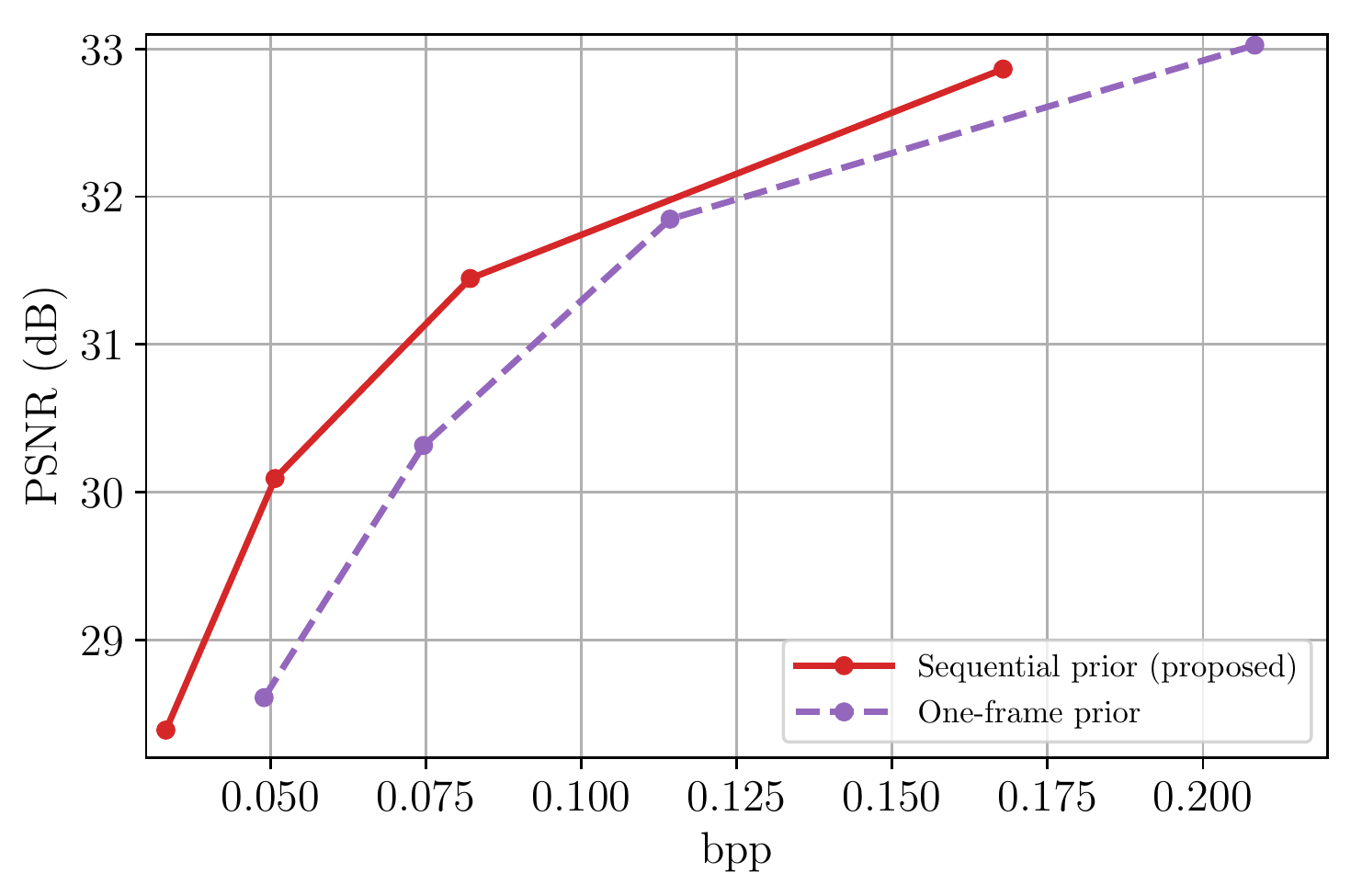}
\caption{The average performance of the 6-th P-frame in all GOPs of the JCT-VC dataset. The two curves correspond to Fig.~\ref{fig:prior} (a) and (b), respectively.}\label{fig:abl_lstm}
\end{figure}

\textbf{The proposed RPM.} It can be seen from Fig.~\ref{fig:abl} that the proposed RPM (BL+RAE+RPM) significantly reduces the bit-rate in comparison with BL+RAE, which uses the spatial entropy model~\cite{balle2017end}. This proves the fact that at the same compression quality, the temporally \textit{conditional} cross entropy is smaller than the \textit{independent} cross entropy, \ie,
\begin{equation}\label{conclusion}\nonumber
\mathbb{E}_{\bm y_t \sim p_t}[-\log_2 q_t(\bm y_t\, | \,\bm y_1, \dots, \bm y_{t-1})] 
< \mathbb{E}_{\bm y_t \sim p}[-\log_2 q(\bm y_t)].
\end{equation}
Besides, Fig.~\ref{fig:abl} shows that our RPM network further outperforms the hyperprior spatial entropy model~\cite{balle2018variational}, which generates the side information $\bm z_t$ to 
facilitate the compression of $\bm y_t$. This indicates that when compressing video at the same quality, the \textit{temporally} conditional cross entropy is smaller than the \textit{spatial} conditional cross entropy (with the overhead cross entropy of $\bm z_t$), \ie,
\begin{equation}\label{conclusion_hyper}\nonumber
\begin{aligned}
&\mathbb{E}_{\bm y_t \sim p_t}[-\log_2 q_t(\bm y_t\, | \,\bm y_1, \dots, \bm y_{t-1})] \\
&< \mathbb{E}_{\bm y_t \sim p_{\bm y|\bm z}}[-\log_2 q_{\bm y|\bm z}(\bm y_t\, | \, \bm z_t)] + \mathbb{E}_{\bm z_t \sim p_{\bm z}}[-\log_2 q_{\bm z}(\bm z_t)].
\end{aligned}
\end{equation}
The proposed RPM has two benefits over \cite{balle2018variational}. First, our RPM does not consume overhead bit-rate to compress the prior information, while \cite{balle2018variational} has to compress $\bm z_t$ into bit stream. Second, our RPM uses the temporal prior of all previous latent representations, while there is only one spatial prior 
$\bm z_t$ in \cite{balle2018variational} with much smaller size, \ie, $\frac{1}{16}$ of $\bm y_t$. In conclusion, these studies verify the benefits of applying temporal prior to estimate the conditional probability $q_{t}(\bm y_{t}\, | \, \bm y_1, \dots, \bm y_{t-1} )$ in a recurrent manner.

\begin{figure*}[!t]
\centering
\subfigure[Change of PSNR (dB) and bpp at $\lambda=512$]{\includegraphics[width=.4\linewidth]{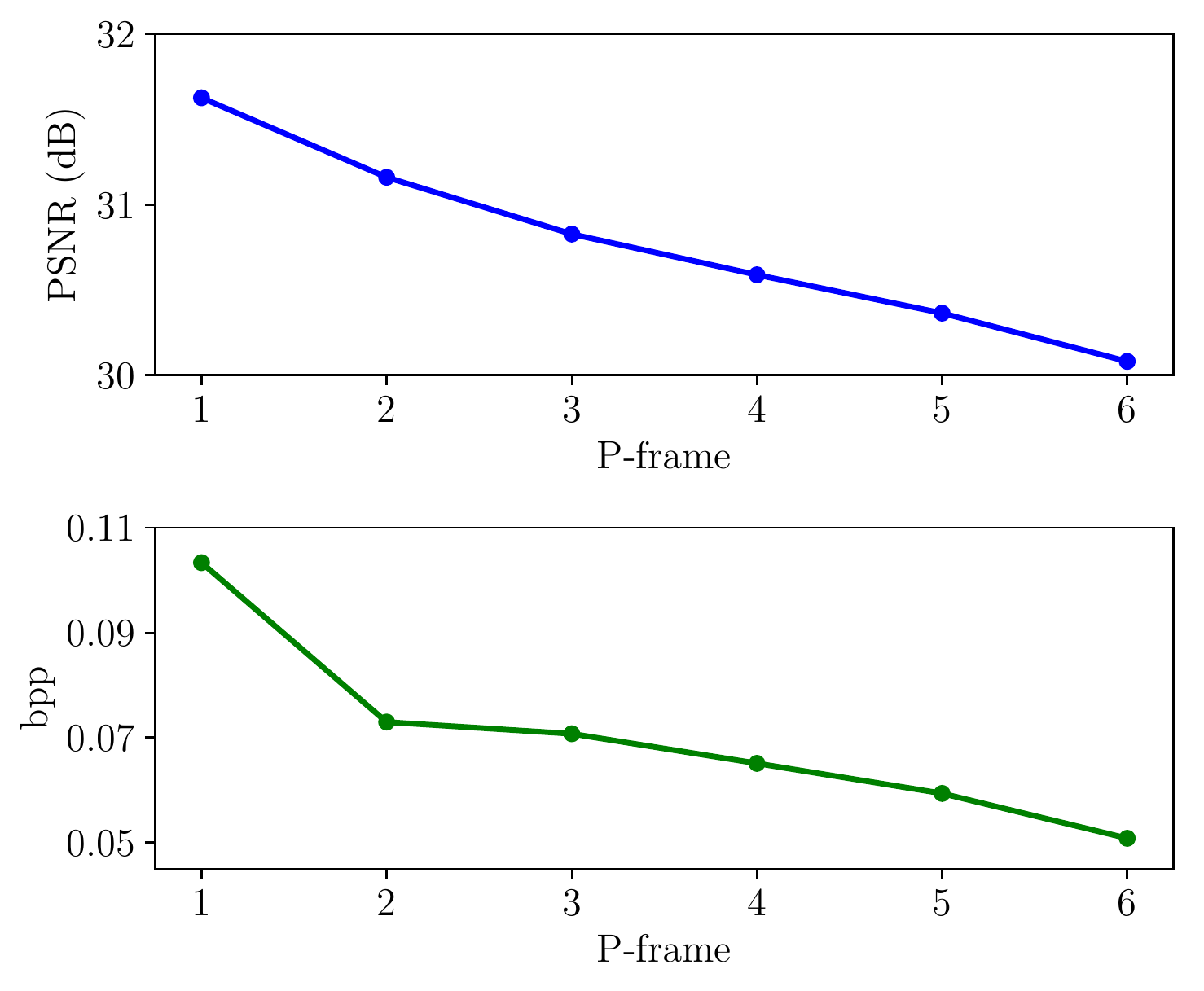}}
\subfigure[Change of PSNR (dB) and bpp at $\lambda=1024$]{\includegraphics[width=.4\linewidth]{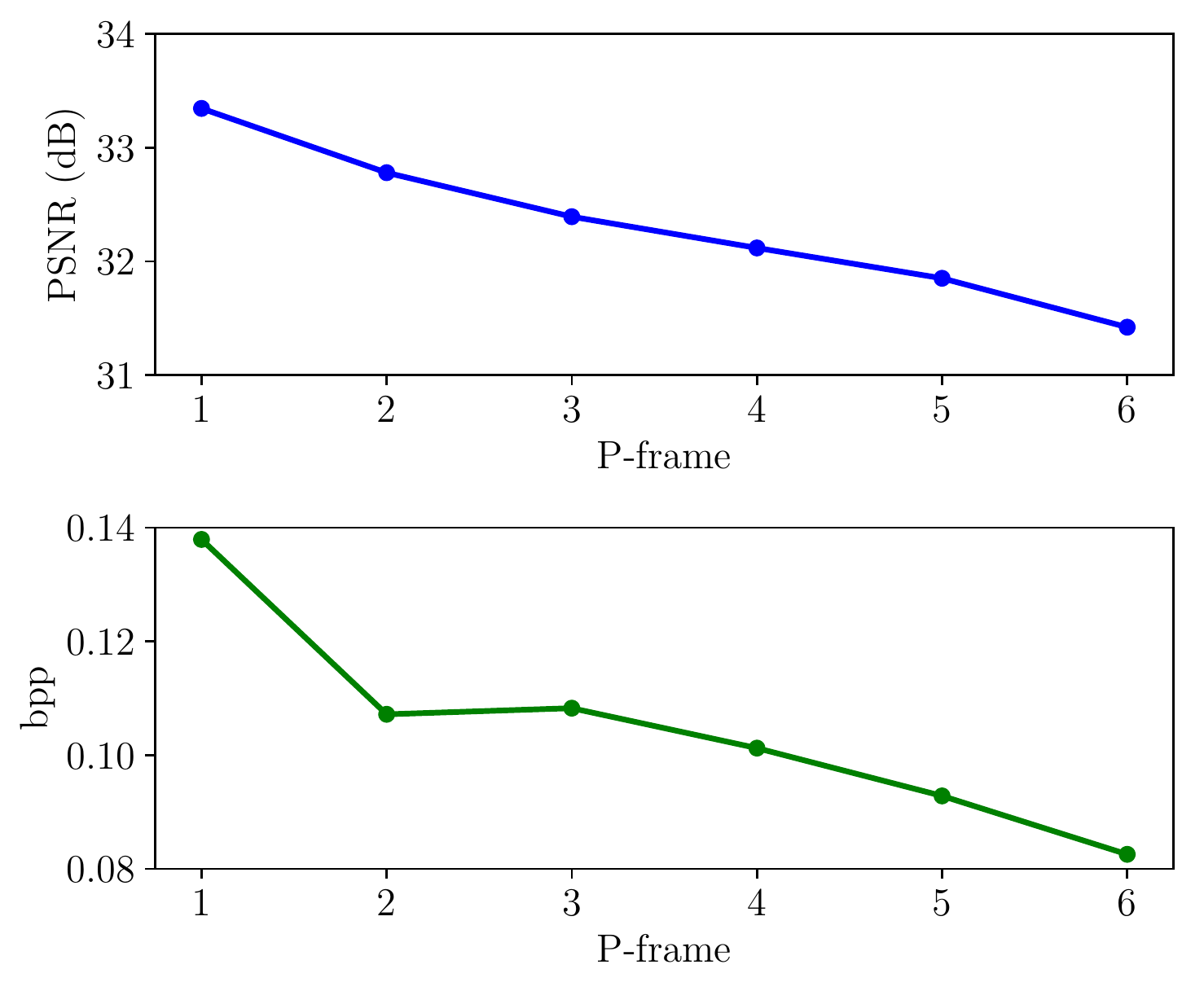}}
\caption{The changes of PSNR (dB) and bpp on consecutive P-frames at (a) $\lambda=512$ and (b) $\lambda=1024$.}\label{fig:stat}
\end{figure*}

\textbf{Sequential prior vs. one-frame prior.} Fig.~\ref{fig:prior} (a) shows the recurrent framework of the proposed RLVC, in which the hidden states are transferred through P-frames. This way, the sequential prior can be utilized to compress the upcoming P-frame, \eg, the 6-th P-frame. To show the advantage of applying recurrent cells in RLVC, instead of only using the prior of the intermediately previous frame, we conduct the ablation to reset the hidden states to the initial state before the 5-th P-frame (as shown in Fig.~\ref{fig:prior} (b)) and analyze the decrease of compression performance on the 6-th frame. Fig.~\ref{fig:abl_lstm} illustrates the performance of the 6-th frame in all Group of Pictures (GOPs) of all videos in the JCT-VC test set. It can be seen from Fig.~\ref{fig:abl_lstm} that only utilizing the one-frame prior, \ie, resetting states before the 5-th P-frame, obviously decreases the rate-distortion performance of the 6-th P-frame, in comparison with using the sequential prior in our RLVC approach. This validates the effectiveness and advantage of the recurrent framework of RLVC.

\textbf{Change of PSNR and bpp along consecutive P-frames.} 
Fig.~\ref{fig:stat} shows the change of PSNR (dB) and bit-rate along sequential P-frames after the I-frame. The results are averaged among all GOPs in the JCT-VC test set. It can be seen from Fig.~\ref{fig:stat} that both PSNR and bit-rate decrease when the distance from I-frame increases. This is probably because of the combined effect of the richer prior transferred in recurrent networks and the farther distance from I-frame. In terms of PSNR, the decrease of quality on the frame which is used to predict the next frame by motion compensation (refer to Fig.~\ref{fig:framework}) may lead to quality decrease of the next frame. However, given more previous frames, the proposed Recurrent Auto-Encoder (RAE) and Recurrent Probability Model (RPM) are both with richer temporal prior. Therefore, the RAE learns to generate more efficient latent representation and the RPM learns to more accurately model the conditional probability function. This way, the bit-rate drops along sequential P-frames.

\subsection{Combining RPM with spatial probability models} 
\begin{figure}[!t]
\includegraphics[width=1\linewidth]{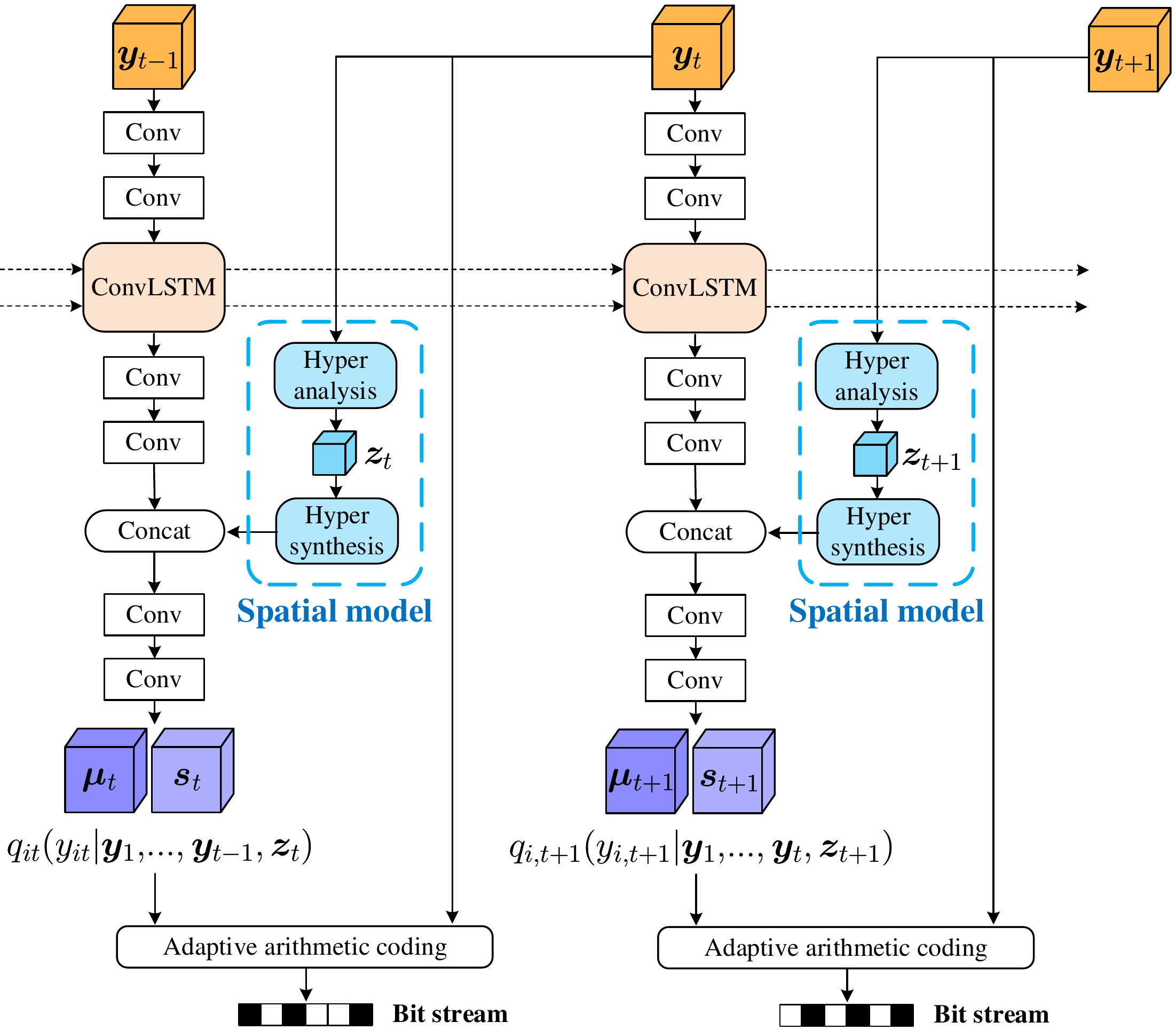}
\caption{The probability model combining the proposed RPM  with the spatial hyperprior model~\cite{balle2018variational}.}\label{fig:hyper}
\end{figure}

It is worth pointing out that the proposed RPM network is flexible to be combined with various spatial probability models, \eg, \cite{balle2018variational,lee2019context,Hu2020Coarse}. As an example, we train a model combining the proposed approach with the hyperprior spatial probability model~\cite{balle2018variational}, which is illustrated in Fig.~\ref{fig:hyper}. This combined model only slightly improves our approach, \ie, bit-rate reduction $<1\%$. 
On the one hand, such slight improvement indicates that due to the high correlation among video frames, the previous latent representations are able to provide most of the useful information, and the spatial prior, which leads to bit-rate overhead, is not very helpful to further improve the performance. This validates the effectiveness of our RPM network. On the other hand, it also shows the flexibility of our RPM network to combine with spatial probability models, \eg, replacing the spatial model in Fig.~\ref{fig:hyper} with \cite{balle2018variational}, \cite{lee2019context} or \cite{Hu2020Coarse}\footnote{Since \cite{lee2019context,Hu2020Coarse} do not release the training codes, we are not able to learn the model combining RPM with \cite{lee2019context,Hu2020Coarse}.}, and the possibility to further advance the performance.

\section{Discussion}\label{discussion}
\subsection{GOP structure}
\begin{figure}[!t]
\centering
\subfigure[The bi-directional IPPP structure (bi-IPPP)]{\includegraphics[width=1\linewidth]{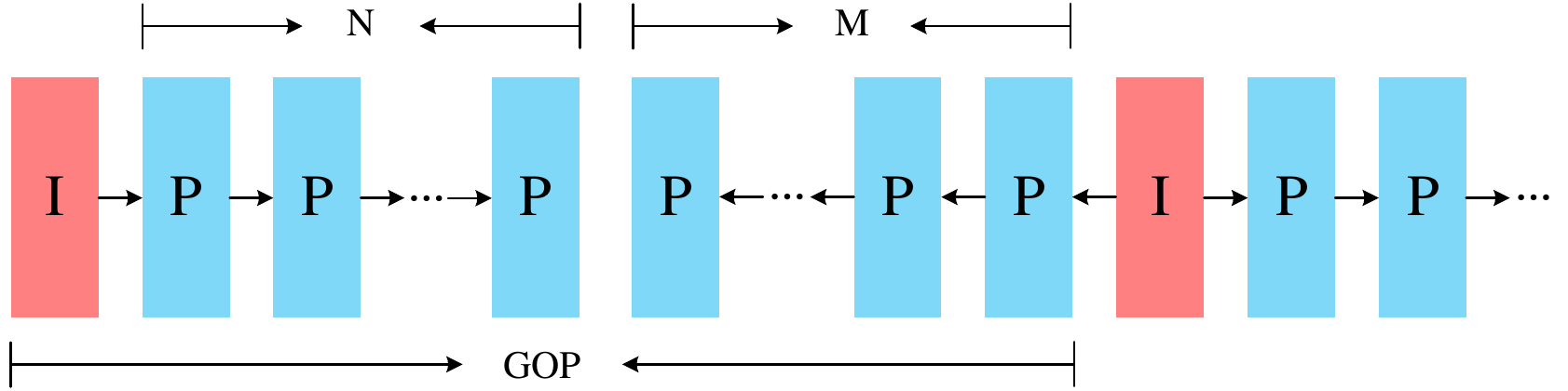}}
\subfigure[The uni-directional IPPP structure (uni-IPPP)]{\includegraphics[width=1\linewidth]{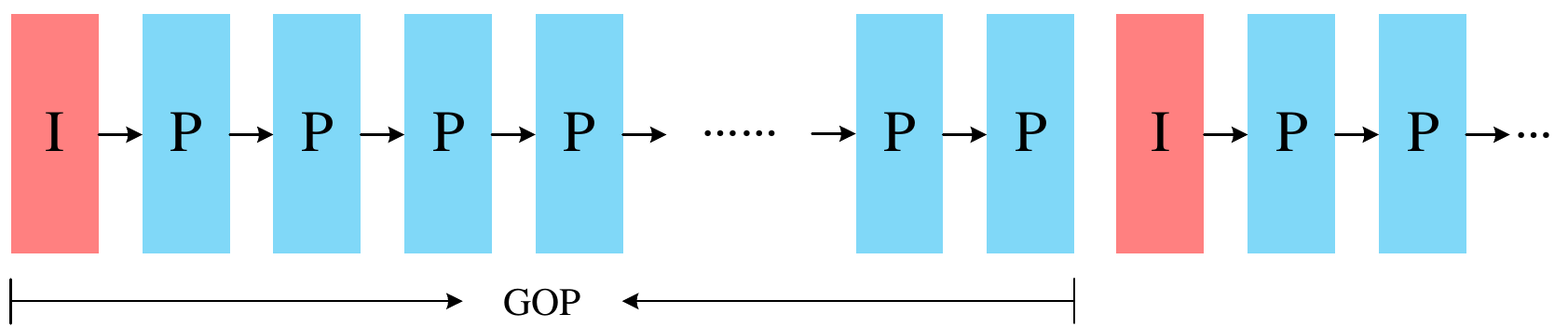}}
\caption{Two kinds of GOP structures for our RLVC approach.}\label{fig:gop}
\end{figure}

In this section, we first discuss the performance of our approach when compressing video with different Group of Picture (GOP) structures. Fig.~\ref{fig:gop} shows two kinds of GOP structures which are possible to be used for the proposed RLVC approach. Fig.~\ref{fig:gop} (a) demonstrates a bi-directional IPPP (bi-IPPP) structure, which reduces the longest distance between I- and P-frames to suppress error propagation. Specifically, $N$ P-frames after the previous I-frame are compressed recurrently by RLVC, then the next I-frame is compressed and the $M$ P-frames before it are recurrently compressed in the reverse direction. This way, the GOP size equals to $N+M+1$. Fig.~\ref{fig:gop} (b) shows the normal IPPP structure (uni-IPPP), which compressed frames in the natural order. 

\begin{figure}[!t]
\centering
\includegraphics[width=.95\linewidth]{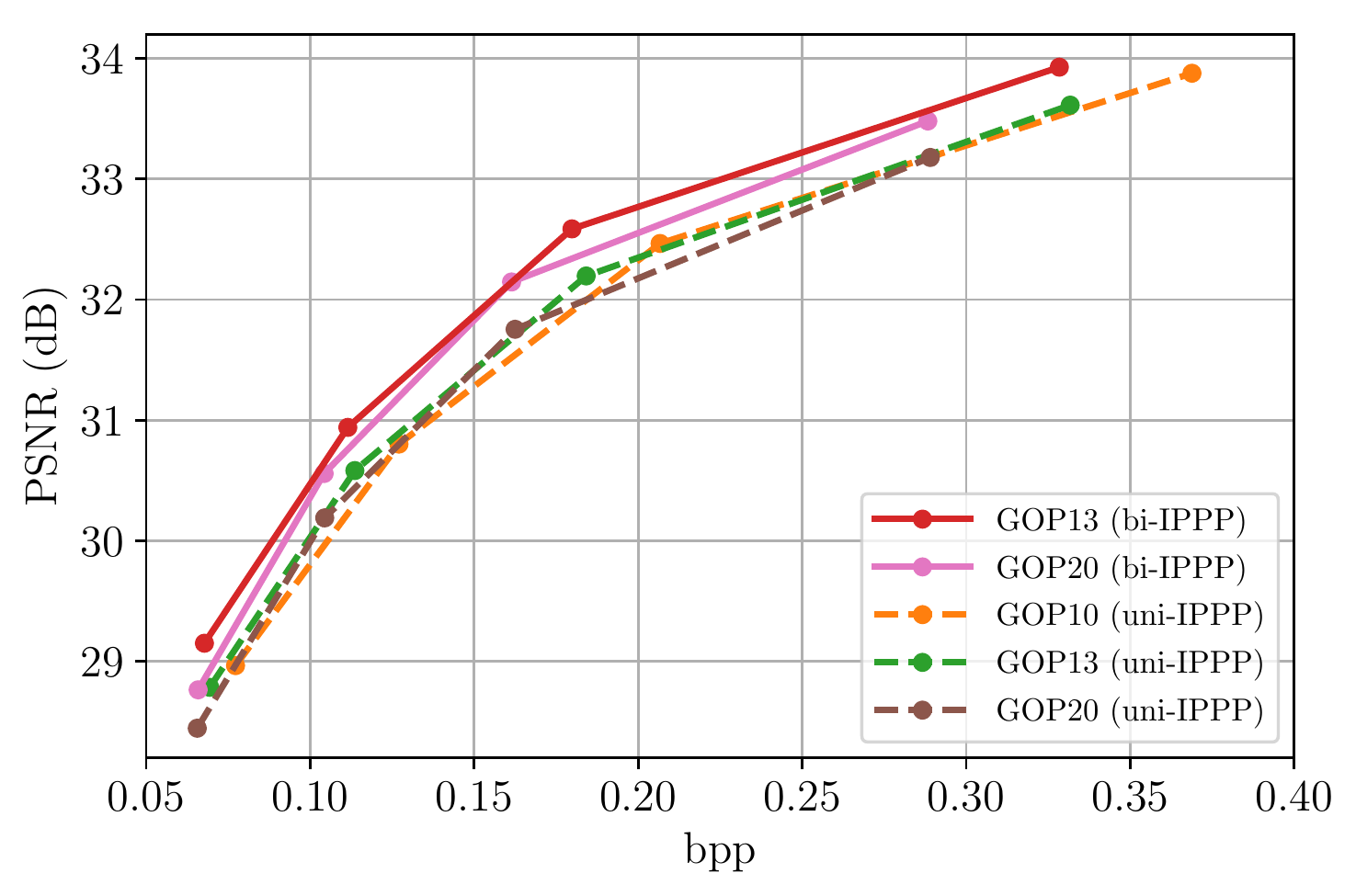}
\caption{The performance of different GOP structures on the JCT-VC dataset.} \label{fig:abl_gop}
\end{figure}

Fig.~\ref{fig:abl_gop} illustrates the rate-distortion curves of our RLVC approach with various GOP structures on the JCT-VC dataset. In Fig.~\ref{fig:abl_gop}, we first show the performance of GOP $=13$ with bi-IPPP mode using $N=M=6$. It achieves the best performance, and this structure is used in the experiments in Section~\ref{exp}. Moreover, Fig.~\ref{fig:abl_gop} shows that when enlarging the GOP size to 20 (bi-IPPP), the performance is still competitive with GOP $=13$ (bi-IPPP) with only slight degradation. Then, we also analyze the uni-IPPP mode (dash lines) from small to large GOP sizes, \ie, GOP $=10, 13$ and $20$. It can be seen from Fig.~\ref{fig:abl_gop} that the performance of the uni-IPPP mode are very similar among different GOP sizes, which are lower than the bi-IPPP mode by around 0.3 dB to 0.5 dB. Note that it also happens to other traditional and learned video compression approaches that bi-directional prediction achieves better performance than the uni-directional prediction. Also, the bi-directional prediction is utilized in previous learned video compression approaches, \eg, Wu~\etal~\cite{wu2018video} and HLVC~\cite{yang2020heirarchical}, and we outperform \cite{wu2018video, yang2020heirarchical} as the results shown in Fig.~\ref{fig:curve}.

In conclusion, the proposed RLVC approach is compatible to various GOP sizes, and especially adjustable to GOP $=20$ (bi-IPPP) without obvious degradation of compression performance.

\subsection{Error propagation}

\begin{figure}[!t]
\centering
\includegraphics[width=.95\linewidth]{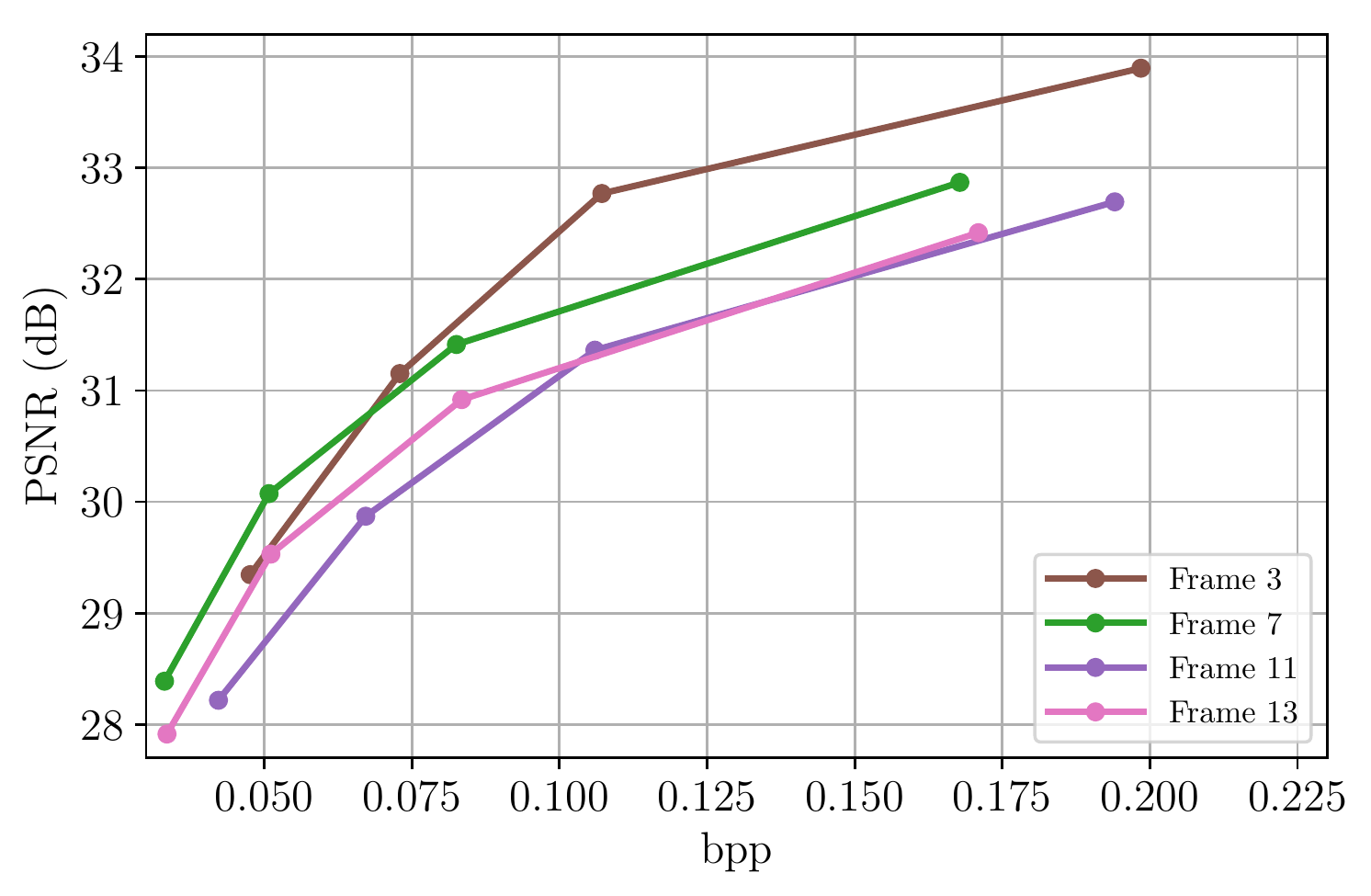}
\caption{The average performance of different P-frames among all GOPs in the JCT-VC dataset.} \label{fig:abl_p}
\end{figure}

Learned video compression approaches usually suffer from error propagation when the distance between I- and P-frames increases. We analyzed the rate-distortion curves of 12 consecutive P-frames compressed by the proposed RLVC model. Note that the length of 12 is twice as long as the training samples (6 P-frames plus an I-frame). Fig.~\ref{fig:abl_p} shows the average performance of different frames with the GOP = 13 (uni-IPPP) setting. The rate-distortion curves are averaged among all GOPs in the JCT-VC dataset. In Fig.~\ref{fig:abl_p}, frame 7 (the 6-th P-frame) and its previous frames are within the training length of our models. It can be seen that after going out of the training length, frames 11 and 13 indeed have lower performance than frame 7, \eg, the PSNR drops around 0.5 dB from frame 7 to frame 13. This indicates that error propagation also exists in the proposed RLVC approach. However, the performance on frame 13 is even better than frame 11 at low bit-rates and maintains comparable performance with frame 11 at high bit-rates. A similar phenomenon can be observed from frame 3 to frame 7, \ie, frame 7 achieves higher performance at low bit-rates than frame 3. This is probably because the proposed recurrent compression network contains richer temporal prior when moving forwards frame-by-frame, thus facilitating the compression of the frames which are farther from I-frame. To a certain degree, this mechanism is able to combat against the error propagation caused by the increasing distance from I-frame. Therefore, as Fig.~\ref{fig:abl_gop} shows, the compression performance does not degrade obviously when enlarging the GOP size.

\section{Conclusion and future work}\label{conc}

This paper has proposed a recurrent learned video compression approach. Specifically, we proposed recurrent auto-encoders to compress motion and residual, fully exploring the temporal correlation in video frames. Then, we showed how modeling the conditional probability in a recurrent manner improves the coding efficiency. The proposed recurrent auto-encoders and recurrent probability model significantly expands the range of reference frames, which has not been achieved in previous learned \textit{as well as} handcrafted standards. 
The experiments validate that the proposed approach outperforms all previous learned approaches and the LDP default setting of x265 in terms of both PSNR and MS-SSIM, and also outperforms x265 (slowest) on MS-SSIM. The ablation studies verify the effectiveness of each recurrent component in our RLVC approach, and show the flexibility of the proposed RPM network to combine with spatial probability models.

Moreover, the proposed method can inspire traditional codecs, particularly the methods that integrate deep networks in traditional codecs, to adopt recurrent networks to improve their performance. For instance, Liu~\etal~\cite{liu2018one} and Choi~\etal~\cite{choi2019deep} improves the motion compensation of HEVC by utilizing single deep network on each frame, and Li~\etal~\cite{li2019densenet, li2019deep} replace the in-loop of HEVC by non-recurrent deep networks. These methods are possible to be advanced by employing the recurrent networks (similar to the proposed approach) to further improve the traditional codecs, such as HEVC.

In this paper, 
the recurrent framework of the proposed approach still relies on the warping operation and motion compensation to reduce the temporal redundancy. Therefore, it is a promising future work to eliminate the dependency on optical-flow-based motion detection, and learn a fully recurrent network or adopt an attention mechanism (\eg, transformer~\cite{vaswani2017attention} based) for learned video compression. Besides, the proposed approach achieves superior performance at the cost of higher encoding complexity. Another future work is to study reducing complexity and the trade-off between complexity and rate-distortion performance. For example, the proposed network may be sped up by reducing the layer number and channel numbers in the auto-encoders and the motion compensation network, or by utilizing a more time-efficient optical flow network for motion prediction.

\bibliographystyle{IEEEtran}
\bibliography{egbib}

\begin{thebibliography}{10}
\providecommand{\url}[1]{#1}
\csname url@samestyle\endcsname
\providecommand{\newblock}{\relax}
\providecommand{\bibinfo}[2]{#2}
\providecommand{\BIBentrySTDinterwordspacing}{\spaceskip=0pt\relax}
\providecommand{\BIBentryALTinterwordstretchfactor}{4}
\providecommand{\BIBentryALTinterwordspacing}{\spaceskip=\fontdimen2\font plus
\BIBentryALTinterwordstretchfactor\fontdimen3\font minus
  \fontdimen4\font\relax}
\providecommand{\BIBforeignlanguage}[2]{{%
\expandafter\ifx\csname l@#1\endcsname\relax
\typeout{** WARNING: IEEEtran.bst: No hyphenation pattern has been}%
\typeout{** loaded for the language `#1'. Using the pattern for}%
\typeout{** the default language instead.}%
\else
\language=\csname l@#1\endcsname
\fi
#2}}
\providecommand{\BIBdecl}{\relax}
\BIBdecl

\bibitem{Cisco}
Cisco, ``Cisco visual networking index: Global mobile data traffic forecast
  update, 2017-2022 white paper,''
  \url{https://www.cisco.com/c/en/us/solutions/collateral/service-provider/visual-networking-index-vni/white-paper-c11-738429.html}.

\bibitem{le1992mpeg}
D.~J. Le~Gall, ``The {MPEG} video compression algorithm,'' \emph{Signal
  Processing: Image Communication}, vol.~4, no.~2, pp. 129--140, 1992.

\bibitem{wiegand2003overview}
T.~Wiegand, G.~J. Sullivan, G.~Bjontegaard, and A.~Luthra, ``Overview of the
  {H.264/AVC} video coding standard,'' \emph{IEEE Transactions on circuits and
  systems for video technology}, vol.~13, no.~7, pp. 560--576, 2003.

\bibitem{sullivan2012overview}
G.~J. Sullivan, J.-R. Ohm, W.-J. Han, and T.~Wiegand, ``Overview of the high
  efficiency video coding ({HEVC}) standard,'' \emph{IEEE Transactions on
  circuits and systems for video technology}, vol.~22, no.~12, pp. 1649--1668,
  2012.

\bibitem{minnen2018joint}
D.~Minnen, J.~Ball{\'e}, and G.~D. Toderici, ``Joint autoregressive and
  hierarchical priors for learned image compression,'' in \emph{Advances in
  Neural Information Processing Systems (NeurIPS)}, 2018, pp. 10\,771--10\,780.

\bibitem{lee2019context}
J.~Lee, S.~Cho, and S.-K. Beack, ``Context-adaptive entropy model for
  end-to-end optimized image compression,'' in \emph{Proceedings of the
  International Conference on Learning Representations (ICLR)}, 2019.

\bibitem{Hu2020Coarse}
Y.~Hu, W.~Yang, and J.~Liu, ``Coarse-to-fine hyper-prior modeling for learned
  image compression,'' in \emph{Proceedings of the AAAI Conference on
  Artificial Intelligence}, 2020.

\bibitem{wu2018video}
C.-Y. Wu, N.~Singhal, and P.~Krahenbuhl, ``Video compression through image
  interpolation,'' in \emph{Proceedings of the European Conference on Computer
  Vision (ECCV)}, 2018, pp. 416--431.

\bibitem{lu2019dvc}
G.~Lu, W.~Ouyang, D.~Xu, X.~Zhang, C.~Cai, and Z.~Gao, ``{DVC}: An end-to-end
  deep video compression framework,'' in \emph{Proceedings of the IEEE
  Conference on Computer Vision and Pattern Recognition (CVPR)}, 2019, pp.
  11\,006--11\,015.

\bibitem{cheng2019learning}
Z.~Cheng, H.~Sun, M.~Takeuchi, and J.~Katto, ``Learning image and video
  compression through spatial-temporal energy compaction,'' in
  \emph{Proceedings of the IEEE Conference on Computer Vision and Pattern
  Recognition (CVPR)}, 2019, pp. 10\,071--10\,080.

\bibitem{djelouah2019neural}
A.~Djelouah, J.~Campos, S.~Schaub-Meyer, and C.~Schroers, ``Neural inter-frame
  compression for video coding,'' in \emph{Proceedings of the IEEE
  International Conference on Computer Vision (ICCV)}, 2019, pp. 6421--6429.

\bibitem{yang2020heirarchical}
R.~Yang, F.~Mentzer, L.~Van~Gool, and R.~Timofte, ``Learning for video
  compression with hierarchical quality and recurrent enhancement,'' in
  \emph{Proceedings of the IEEE Conference on Computer Vision and Pattern
  Recognition (CVPR)}, 2020.

\bibitem{hinton1994autoencoders}
G.~E. Hinton and R.~S. Zemel, ``Autoencoders, minimum description length and
  helmholtz free energy,'' in \emph{Advances in neural information processing
  systems}, 1994, pp. 3--10.

\bibitem{cho2013simple}
K.~Cho, ``Simple sparsification improves sparse denoising autoencoders in
  denoising highly corrupted images,'' in \emph{Proceedings of the
  International Conference on Machine Learning (ICML)}, 2013, pp. 432--440.

\bibitem{gondara2016medical}
L.~Gondara, ``Medical image denoising using convolutional denoising
  autoencoders,'' in \emph{2016 IEEE 16th International Conference on Data
  Mining Workshops (ICDMW)}.\hskip 1em plus 0.5em minus 0.4em\relax IEEE, 2016,
  pp. 241--246.

\bibitem{lore2017llnet}
K.~G. Lore, A.~Akintayo, and S.~Sarkar, ``Llnet: A deep autoencoder approach to
  natural low-light image enhancement,'' \emph{Pattern Recognition}, vol.~61,
  pp. 650--662, 2017.

\bibitem{park2018dual}
S.~Park, S.~Yu, M.~Kim, K.~Park, and J.~Paik, ``Dual autoencoder network for
  retinex-based low-light image enhancement,'' \emph{IEEE Access}, vol.~6, pp.
  22\,084--22\,093, 2018.

\bibitem{zeng2015coupled}
K.~Zeng, J.~Yu, R.~Wang, C.~Li, and D.~Tao, ``Coupled deep autoencoder for
  single image super-resolution,'' \emph{IEEE transactions on cybernetics},
  vol.~47, no.~1, pp. 27--37, 2015.

\bibitem{wang2016non}
R.~Wang and D.~Tao, ``Non-local auto-encoder with collaborative stabilization
  for image restoration,'' \emph{IEEE Transactions on Image Processing},
  vol.~25, no.~5, pp. 2117--2129, 2016.

\bibitem{karpathy2015visualizing}
A.~Karpathy, J.~Johnson, and L.~Fei-Fei, ``Visualizing and understanding
  recurrent networks,'' 2016.

\bibitem{mikolov2010recurrent}
T.~Mikolov, M.~Karafi{\'a}t, L.~Burget, J.~{\v{C}}ernock{\`y}, and
  S.~Khudanpur, ``Recurrent neural network based language model,'' in
  \emph{Eleventh annual conference of the international speech communication
  association}, 2010.

\bibitem{jozefowicz2016exploring}
R.~Jozefowicz, O.~Vinyals, M.~Schuster, N.~Shazeer, and Y.~Wu, ``Exploring the
  limits of language modeling,'' \emph{arXiv preprint arXiv:1602.02410}, 2016.

\bibitem{donahue2015long}
J.~Donahue, L.~Anne~Hendricks, S.~Guadarrama, M.~Rohrbach, S.~Venugopalan,
  K.~Saenko, and T.~Darrell, ``Long-term recurrent convolutional networks for
  visual recognition and description,'' in \emph{Proceedings of the IEEE
  Conference on Computer Vision and Pattern Recognition (CVPR)}, 2015, pp.
  2625--2634.

\bibitem{cho2014learning}
K.~Cho, B.~van Merri{\"e}nboer, C.~Gulcehre, D.~Bahdanau, F.~Bougares,
  H.~Schwenk, and Y.~Bengio, ``Learning phrase representations using rnn
  encoder--decoder for statistical machine translation,'' in \emph{Proceedings
  of the 2014 Conference on Empirical Methods in Natural Language Processing
  (EMNLP)}, 2014, pp. 1724--1734.

\bibitem{sutskever2014sequence}
I.~Sutskever, O.~Vinyals, and Q.~V. Le, ``Sequence to sequence learning with
  neural networks,'' in \emph{Advances in neural information processing
  systems}, 2014, pp. 3104--3112.

\bibitem{vinyals2015show}
O.~Vinyals, A.~Toshev, S.~Bengio, and D.~Erhan, ``Show and tell: A neural image
  caption generator,'' in \emph{Proceedings of the IEEE conference on computer
  vision and pattern recognition (CVPR)}, 2015, pp. 3156--3164.

\bibitem{srivastava2015unsupervised}
N.~Srivastava, E.~Mansimov, and R.~Salakhudinov, ``Unsupervised learning of
  video representations using {LSTM}s,'' in \emph{Proceedings of the
  International Conference on Machine Learning (ICML)}, 2015, pp. 843--852.

\bibitem{hochreiter1997long}
S.~Hochreiter and J.~Schmidhuber, ``Long short-term memory,'' \emph{Neural
  computation}, vol.~9, no.~8, pp. 1735--1780, 1997.

\bibitem{Toderici2016Variable}
G.~Toderici, S.~M. O'Malley, S.~J. Hwang, D.~Vincent, D.~Minnen, S.~Baluja,
  M.~Covell, and R.~Sukthankar, ``Variable rate image compression with
  recurrent neural networks,'' in \emph{Proceedings of the International
  Conference on Learning Representations (ICLR)}, 2016.

\bibitem{toderici2017full}
G.~Toderici, D.~Vincent, N.~Johnston, S.~Jin~Hwang, D.~Minnen, J.~Shor, and
  M.~Covell, ``Full resolution image compression with recurrent neural
  networks,'' in \emph{Proceedings of the IEEE Conference on Computer Vision
  and Pattern Recognition (CVPR)}, 2017, pp. 5306--5314.

\bibitem{agustsson2017soft}
E.~Agustsson, F.~Mentzer, M.~Tschannen, L.~Cavigelli, R.~Timofte, L.~Benini,
  and L.~V. Gool, ``Soft-to-hard vector quantization for end-to-end learning
  compressible representations,'' in \emph{Advances in Neural Information
  Processing Systems (NeurIPS)}, 2017, pp. 1141--1151.

\bibitem{theis2017lossy}
L.~Theis, W.~Shi, A.~Cunningham, and F.~Husz{\'a}r, ``Lossy image compression
  with compressive autoencoders,'' in \emph{Proceedings of the International
  Conference on Learning Representations (ICLR)}, 2017.

\bibitem{balle2017end}
J.~Ball{\'e}, V.~Laparra, and E.~P. Simoncelli, ``End-to-end optimized image
  compression,'' in \emph{Proceedings of the International Conference on
  Learning Representations (ICLR)}, 2017.

\bibitem{balle2018variational}
J.~Ball{\'e}, D.~Minnen, S.~Singh, S.~J. Hwang, and N.~Johnston, ``Variational
  image compression with a scale hyperprior,'' in \emph{Proceedings of the
  International Conference on Learning Representations (ICLR)}, 2018.

\bibitem{mentzer2018conditional}
F.~Mentzer, E.~Agustsson, M.~Tschannen, R.~Timofte, and L.~Van~Gool,
  ``Conditional probability models for deep image compression,'' in
  \emph{Proceedings of the IEEE Conference on Computer Vision and Pattern
  Recognition (CVPR)}, 2018, pp. 4394--4402.

\bibitem{li2018learning}
M.~Li, W.~Zuo, S.~Gu, D.~Zhao, and D.~Zhang, ``Learning convolutional networks
  for content-weighted image compression,'' in \emph{Proceedings of the IEEE
  Conference on Computer Vision and Pattern Recognition (CVPR)}, 2018, pp.
  3214--3223.

\bibitem{johnston2018improved}
N.~Johnston, D.~Vincent, D.~Minnen, M.~Covell, S.~Singh, T.~Chinen,
  S.~Jin~Hwang, J.~Shor, and G.~Toderici, ``Improved lossy image compression
  with priming and spatially adaptive bit rates for recurrent networks,'' in
  \emph{Proceedings of the IEEE Conference on Computer Vision and Pattern
  Recognition (CVPR)}, 2018, pp. 4385--4393.

\bibitem{skodras2001jpeg}
A.~Skodras, C.~Christopoulos, and T.~Ebrahimi, ``The {JPEG} 2000 still image
  compression standard,'' \emph{IEEE Signal processing magazine}, vol.~18,
  no.~5, pp. 36--58, 2001.

\bibitem{BPG}
F.~Bellard, ``{BPG} image format,'' \url{https://bellard.org/bpg/}.

\bibitem{xu2018reducing}
M.~Xu, T.~Li, Z.~Wang, X.~Deng, R.~Yang, and Z.~Guan, ``Reducing complexity of
  {HEVC}: A deep learning approach,'' \emph{IEEE Transactions on Image
  Processing}, vol.~27, no.~10, pp. 5044--5059, 2018.

\bibitem{liu2018one}
J.~Liu, S.~Xia, W.~Yang, M.~Li, and D.~Liu, ``One-for-all: Grouped variation
  network-based fractional interpolation in video coding,'' \emph{IEEE
  Transactions on Image Processing}, vol.~28, no.~5, pp. 2140--2151, 2018.

\bibitem{choi2019deep}
H.~Choi and I.~V. Baji{\'c}, ``Deep frame prediction for video coding,''
  \emph{IEEE Transactions on Circuits and Systems for Video Technology}, 2019.

\bibitem{dai2017convolutional}
Y.~Dai, D.~Liu, and F.~Wu, ``A convolutional neural network approach for
  post-processing in {HEVC} intra coding,'' in \emph{Proceedings of the
  International Conference on Multimedia Modeling (MMM)}.\hskip 1em plus 0.5em
  minus 0.4em\relax Springer, 2017, pp. 28--39.

\bibitem{li2019densenet}
T.~Li, M.~Xu, R.~Yang, and X.~Tao, ``A {DenseNet} based approach for
  multi-frame in-loop filter in {HEVC},'' in \emph{Proceedings of the Data
  Compression Conference (DCC)}.\hskip 1em plus 0.5em minus 0.4em\relax IEEE,
  2019, pp. 270--279.

\bibitem{li2019deep}
T.~Li, M.~Xu, C.~Zhu, R.~Yang, Z.~Wang, and Z.~Guan, ``A deep learning approach
  for multi-frame in-loop filter of {HEVC},'' \emph{IEEE Transactions on Image
  Processing}, 2019.

\bibitem{chen2017deepcoder}
T.~Chen, H.~Liu, Q.~Shen, T.~Yue, X.~Cao, and Z.~Ma, ``Deep{C}oder: A deep
  neural network based video compression,'' in \emph{Proceedings of the IEEE
  Visual Communications and Image Processing (VCIP)}.\hskip 1em plus 0.5em
  minus 0.4em\relax IEEE, 2017, pp. 1--4.

\bibitem{chen2019learning}
Z.~Chen, T.~He, X.~Jin, and F.~Wu, ``Learning for video compression,''
  \emph{IEEE Transactions on Circuits and Systems for Video Technology}, 2019.

\bibitem{habibian2019video}
A.~Habibian, T.~van Rozendaal, J.~M. Tomczak, and T.~S. Cohen, ``Video
  compression with rate-distortion autoencoders,'' in \emph{Proceedings of the
  IEEE International Conference of Computer Vision (ICCV)}, 2019.

\bibitem{liu2019learned}
H.~Liu, L.~Huang, M.~Lu, T.~Chen, and Z.~Ma, ``Learned video compression via
  joint spatial-temporal correlation exploration,'' in \emph{Proceedings of the
  AAAI Conference on Artificial Intelligence}, 2020.

\bibitem{agustsson2020scale}
E.~Agustsson, D.~Minnen, N.~Johnston, J.~Balle, S.~J. Hwang, and G.~Toderici,
  ``Scale-space flow for end-to-end optimized video compression,'' in
  \emph{Proceedings of the IEEE/CVF Conference on Computer Vision and Pattern
  Recognition (CVPR)}, 2020, pp. 8503--8512.

\bibitem{van2016conditional}
A.~Van~den Oord, N.~Kalchbrenner, L.~Espeholt, O.~Vinyals, A.~Graves
  \emph{et~al.}, ``Conditional image generation with pixelcnn decoders,'' in
  \emph{Advances in neural information processing systems}, 2016, pp.
  4790--4798.

\bibitem{ranjan2017optical}
A.~Ranjan and M.~J. Black, ``Optical flow estimation using a spatial pyramid
  network,'' in \emph{Proceedings of the IEEE Conference on Computer Vision and
  Pattern Recognition (CVPR)}, 2017, pp. 4161--4170.

\bibitem{yang2020opendvc}
R.~Yang, L.~Van~Gool, and R.~Timofte, ``Open{DVC}: An open source
  implementation of the {DVC} video compression method,'' \emph{arXiv preprint
  arXiv:2006.15862}, 2020.

\bibitem{langdon1984introduction}
G.~G. Langdon, ``An introduction to arithmetic coding,'' \emph{IBM Journal of
  Research and Development}, vol.~28, no.~2, pp. 135--149, 1984.

\bibitem{xingjian2015convolutional}
S.~Xingjian, Z.~Chen, H.~Wang, D.-Y. Yeung, W.-K. Wong, and W.-c. Woo,
  ``Convolutional lstm network: A machine learning approach for precipitation
  nowcasting,'' in \emph{Advances in neural information processing systems},
  2015, pp. 802--810.

\bibitem{balakrishnan1991handbook}
N.~Balakrishnan, \emph{Handbook of the logistic distribution}.\hskip 1em plus
  0.5em minus 0.4em\relax CRC Press, 1991.

\bibitem{xue2019video}
T.~Xue, B.~Chen, J.~Wu, D.~Wei, and W.~T. Freeman, ``Video enhancement with
  task-oriented flow,'' \emph{International Journal of Computer Vision}, vol.
  127, no.~8, pp. 1106--1125, 2019.

\bibitem{kingma2014adam}
D.~P. Kingma and J.~Ba, ``Adam: A method for stochastic optimization,'' in
  \emph{Proceedings of the International Conference on Learning Representations
  (ICLR)}, 2015.

\bibitem{bossen2013common}
F.~Bossen, ``Common test conditions and software reference configurations,''
  \emph{JCTVC-L1100}, vol.~12, 2013.

\bibitem{UVG}
A.~Mercat, M.~Viitanen, and J.~Vanne, ``{UVG} dataset: 50/120fps 4{K} sequences
  for video codec analysis and development,'' in \emph{Proceedings of the 11th
  ACM Multimedia Systems Conference}, 2020, pp. 297--302.

\bibitem{wang2016mcl}
H.~Wang, W.~Gan, S.~Hu, J.~Y. Lin, L.~Jin, L.~Song, P.~Wang, I.~Katsavounidis,
  A.~Aaron, and C.-C.~J. Kuo, ``{MCL-JCV}: a {JND}-based {H}.264/{AVC} video
  quality assessment dataset,'' in \emph{Proceedings of the IEEE International
  Conference on Image Processing (ICIP)}.\hskip 1em plus 0.5em minus
  0.4em\relax IEEE, 2016, pp. 1509--1513.

\bibitem{bjontegaard}
G.~Bjontegaard, ``Calculation of average {PSNR} differences between
  {RD}-curves,'' \emph{VCEG-M33}, 2001.

\bibitem{vaswani2017attention}
A.~Vaswani, N.~Shazeer, N.~Parmar, J.~Uszkoreit, L.~Jones, A.~N. Gomez,
  {\L}.~Kaiser, and I.~Polosukhin, ``Attention is all you need,'' in
  \emph{Advances in Neural Information Processing Systems (NeurIPS)}, 2017, pp.
  5998--6008.

\end{thebibliography}

\ifCLASSOPTIONcaptionsoff
  \newpage
\fi

\begin{IEEEbiography}[{\includegraphics[width=1\linewidth]{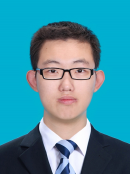}}]{Ren Yang} is a doctoral student at ETH Zurich, Switzerland. He received his M.Sc. degree in 2019 at the School of Electronic and Information Engineering, Beihang University, China, and his B.Sc. degree at the same university in 2016. His research interests include computer vision and video compression. He has published several papers in top international journals and conferences, such as IEEE TPAMI, IEEE TIP, IEEE TCSVT, CVPR, ICCV and ICME. He serves as a reviewer for top conferences and journals, such as CVPR, IJCAI, ECCV, IEEE TIP, IEEE JSTSP, IEEE TCSVT, IEEE TMM, IEEE Access and Elsevier's SPIC and NEUCOM. He got the winner award in the Three Minute Thesis (3MT) Competition at IEEE ICME 2019.
\end{IEEEbiography}

\begin{IEEEbiography}[{\includegraphics[width=1\linewidth]{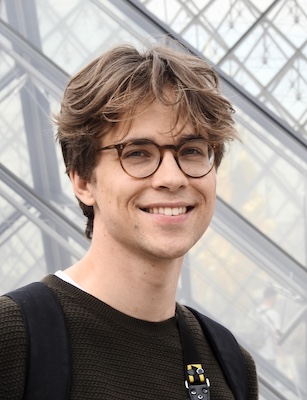}}]{Fabian Mentzer}
is a doctoral student of the Computer Vision Laboratory (CVL) at ETH Zurich, Switzerland. He received his M.Sc. degree in Electrical Engineering and Information Technology from ETH Zurich in 2016, and received his B.Sc. degree from ETH Zurich in 2014. His current research interests include deep learning, learned lossy and lossless image compression, and learned video compression. He has published several papers in top international conferences, such as CVPR, ICCV and NeurIPS. He regularly serves as a reviewer for top conferences such as CVPR, ICCV, ECCV, and NeurIPS. He is a co-organizer of the CLIC workshop at CVPR. 
\end{IEEEbiography}

\begin{IEEEbiography}[{\includegraphics[width=1\linewidth]{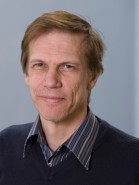}}]{Luc Van Gool}
received the degree in electromechanical engineering at the Katholieke Universiteit Leuven, in 1981. Currently, he is a professor at the Katholieke Universiteit Leuven in Belgium and the ETH in Zurich, Switzerland. He leads computer vision research at both places, where he also teaches computer vision. He has been a program committee member of several major computer vision conferences. His main interests include 3D reconstruction and modeling, object recognition, tracking, and gesture analysis, and the combination of those. He received several Best Paper awards and was nominated Distinguished Researcher by the IEEE Computer Science committee. He is a co-founder of 10 spin-off companies. He is a member of the IEEE.
\end{IEEEbiography}

\begin{IEEEbiography}[{\includegraphics[width=1\linewidth]{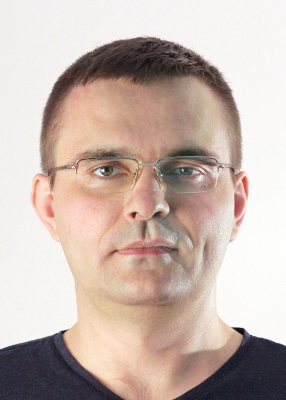}}]{Radu Timofte} 
received his PhD degree in Electrical Engineering from the KU Leuven, in 2013. He is currently a lecturer and a group leader at ETH Zurich, Switzerland. He is a member of the editorial board of top journals such as IEEE TPAMI, Elsevier's CVIU and NEUCOM, and SIAM's SIIMS. He regularly serves as a reviewer and as an area chair for top conferences such as CVPR, ICCV, IJCAI, and ECCV. His work received several awards. He is a co-founder of Merantix and a co-organizer of NTIRE, CLIC, AIM, and PIRM workshops and challenges. His current research interests include deep learning, implicit models, compression, image restoration and enhancement.
\end{IEEEbiography}

\clearpage

\setcounter{page}{1}
\onecolumn
\begin{center}
\LARGE
Learning for Video Compression with Recurrent Auto-Encoder and Recurrent Probability Model
\vspace{.5em}

-- Supporting Document --
\end{center}

\vspace{1em}

\subsection{Performance on conversational video}

\begin{figure}[!h]
\centering
\subfigure{\includegraphics[width=.95\linewidth]{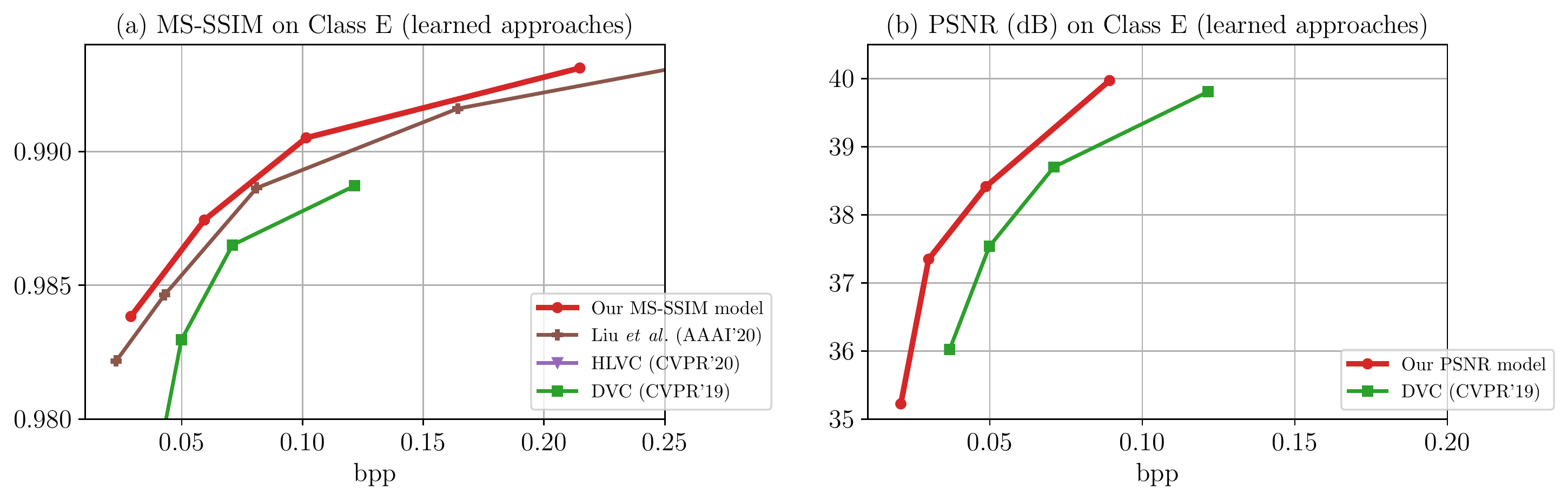}}
\subfigure{\includegraphics[width=.95\linewidth]{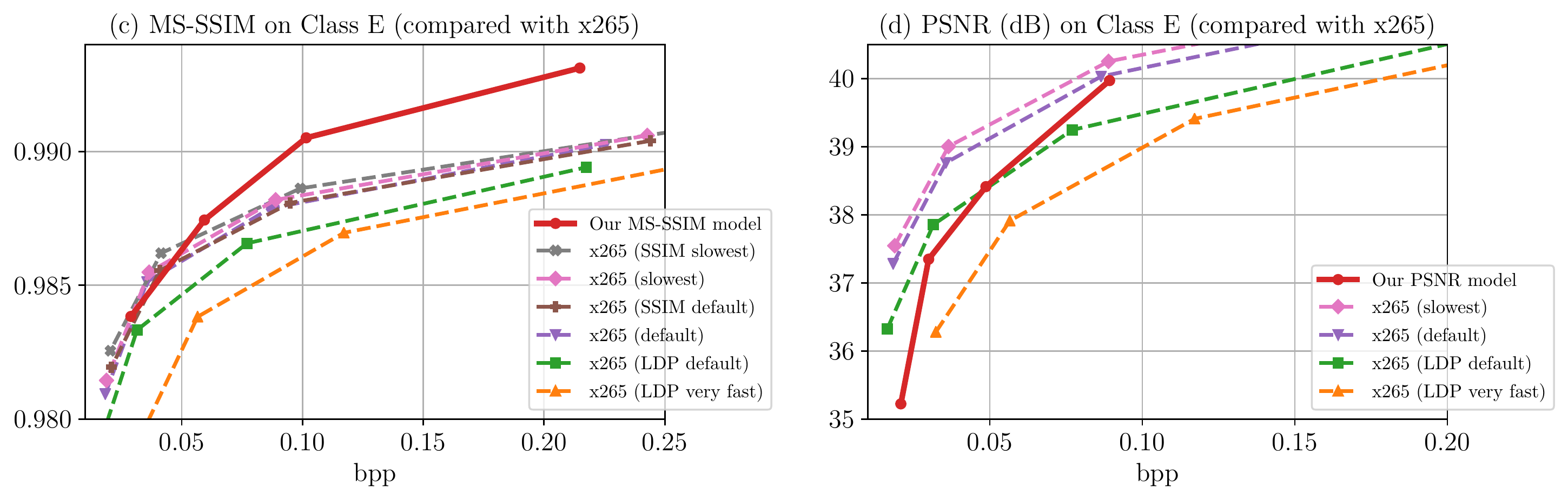}}
\caption{The rate-distortion performance on JCT-VC Class E in comparison with learned approaches and the different settings of x265.} \label{fig:curve_E}
\end{figure}

To validate the generalization ability of the proposed approach, we test our approach on JCT-VC Class E, which is a conversational video dataset. It can be seen from Fig.~\ref{fig:curve_E} (a) and (b) that our RLVC approach outperforms the learned approaches DVC~\cite{lu2019dvc} and Liu~\etal~\cite{liu2019learned} in terms of both MS-SSIM and PSNR.\footnote{Other learned approaches are not tested on JCT-VC Class E, and the MS-SSIM optimized approach Liu~\etal~\cite{liu2019learned} does not have results on PSNR.}  Fig.~\ref{fig:curve_E} (c) shows that our MS-SSIM model outperforms x265 (LDP default) and x265 (LDP very fast) for all bit-rates. We further outperform all other settings (including the SSIM-tuned slowest setting) of x265 at medium and high bit-rates in terms of MS-SSIM, and we are comparable with them at low bit-rates. In terms of PSNR, Fig.~\ref{fig:curve_E} (d) shows that our PSNR model is better than x265 (LDP veryfast), and outperforms x265 (LDP default) when bpp $>$ 0.05.
The same as on UVG and JCT-VC Classes B, C and D, we do not outperform x265 (default) and x265 (slowest) on JCT-VC Class E in terms of PSNR. Recall that, x265 (default) and x265 (slowest) use bi-directional prediction and hierarchical frame structure, but only the uni-directional IPPP mode is applied in our approach. 

\subsection{Performance on the MCL-JCV dataset}

\begin{figure}[!t]
\centering
\includegraphics[width=.95\linewidth]{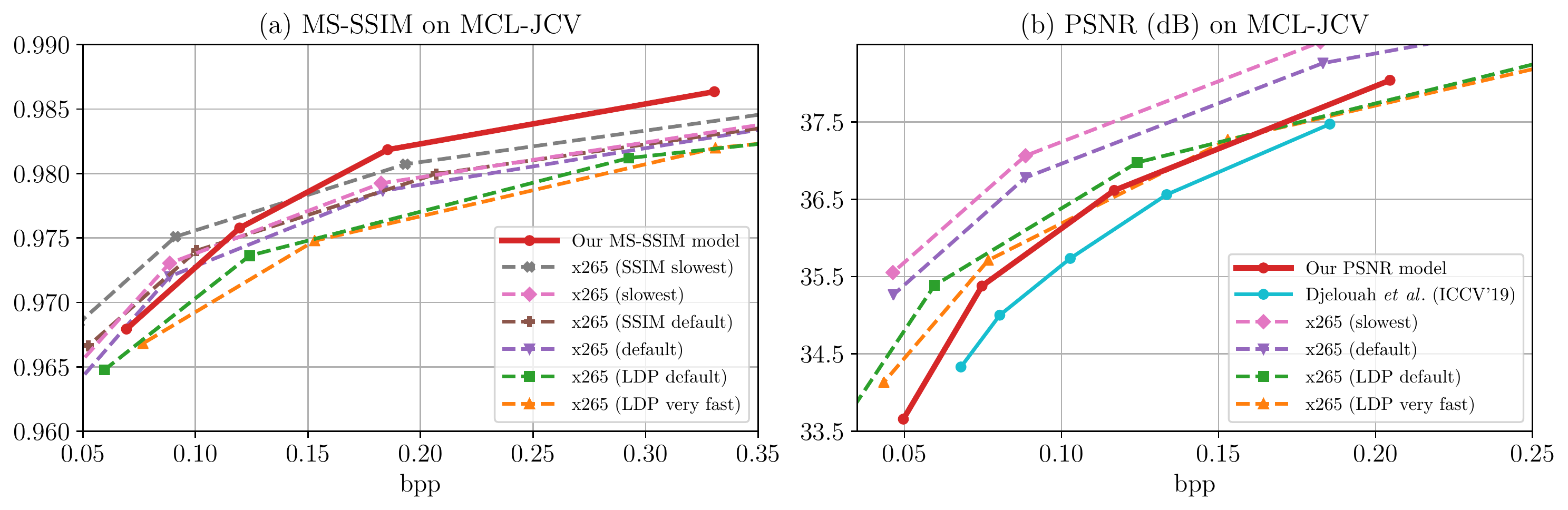}

\caption{The rate-distortion performance on the MCL-JCV dataset. The learned approaches are shown in solid lines and x265 is shown in dash lines.} \label{fig:MCL}
\end{figure}

Fig.~\ref{fig:MCL} demonstrates the rate-distortion performance on the MCL-JCV dataset\footnote{The MCL-JCV dataset is available at \url{http://mcl.usc.edu/mcl-jcv-dataset/.}}, which contains 30 videos with the resolution of $1920\times1080$. As Fig.~\ref{fig:MCL} shows, the proposed MS-SSIM model outperforms the LDP default and the LDP very fast settings of x265, and also outperforms x265 (default), x265 (SSIM default), x265 (slowest) and x265 (SSIM slowest) at high bit-rates. Our MS-SSIM model is comparable with x265 (default) at low bit-rates in terms of MS-SSIM. In terms of PSNR, the proposed PSNR model achieves better performance than the learned video compression approach Djelouah~\etal~\cite{djelouah2019neural} (ICCV'19). 
Note that Djelouah~\etal~\cite{djelouah2019neural} compresses video frame with bi-directional prediction, while the proposed approach only works in the IPPP mode. This proves the superior performance of the proposed recurrent video compression approach. Fig.~\ref{fig:MCL} also indicates that we are comparable with x265 (LDP very fast) on PSNR when bpp $>$ 0.1, and even better than x265 (LDP default) at bpp = 0.2. The same as on other datasets, out PSNR model do not reach better performance than x265 (default) and x265 (slowest), which adopts complicated frame structure and coding strategies.

\end{document}